\documentclass{article}
\pdfoutput=1
\usepackage{arxiv}
\usepackage[utf8x]{inputenc} 
\usepackage[T1]{fontenc}    
\usepackage{caption}
\usepackage{subcaption}
\usepackage{graphicx}
\usepackage{color}
\usepackage[section]{placeins}
\usepackage[normalem]{ulem}
\usepackage{amsmath,amsbsy,amssymb}
\usepackage{hyperref}
\usepackage{tikz}
\usepackage[title]{appendix}
\usepackage[capitalise]{cleveref}
\usepackage{adjustbox}
\usepackage{xcolor}
\definecolor{PPPtext}{rgb}{0.3, 0.0, 0.3}
\definecolor{PPPcomment}{rgb}{0.6, 0.0, 0.0}
\usepackage{subcaption}

\newcommand{\littleo}{o}
\newcommand{\Pdf}{\mathcal{P}}
\newtheorem{lemma}{Lemma}

\begin{document}


\title{A high-order semi-Lagrangian method for  the consistent Monte-Carlo solution
	of stochastic Lagrangian drift-diffusion models coupled with Eulerian discontinuous spectral element method}

\author{Natarajan, H. \textsuperscript{1}, Popov, P.P.   \textsuperscript{1}, Jacobs, G.B.  \textsuperscript{1,*} 
\\ \\
\bigskip	
\textsuperscript{1} Department of Aerospace Engineering, San Diego State University, San Diego, USA \\
{* Corresponding author email address: gjacobs@sdsu.edu}
}

\maketitle

\begin{abstract}
The explicit semi-Lagrangian method method for solution of Lagrangian transport equations
as developed in [Natarajan and Jacobs, \textit{Computer and Fluids}, 2020] is  adopted for the
solution  of stochastic differential
equations that is consistent with Discontinuous Spectral Element Method (DSEM) approximations of Eulerian conservation laws. The method extends the favorable properties of DSEM that include its high-order
accuracy, its local and boundary fitted properties and its high performance on parallel platforms for the concurrent Monte-Carlo, semi-Lagrangian and Eulerian solution of a class of time-dependent problems that can be described by coupled Eulerian-Lagrangian formulations. Such formulations include the probabilistic models used for the simulation of chemically reacting turbulent flows or particle-laden flows.
Consistent with an explicit, DSEM discretization, the semi-Lagrangian method seeds particles at Gauss quadrature collocation nodes within a spectral element. The particles are integrated explicitly in time according to a drift velocity and a Wiener increment forcing and form the nodal basis for an advected interpolant. This interpolant is mapped back in a semi-Lagrangian fashion to the 	Gauss quadrature points through a least squares fit using constraints for
element boundary values. Stochastic Monte-Carlo samples are averaged element-wise on the quadrature nodes. The stable explicit time step
Wiener increment is sufficiently small to prevent 	particles from leaving the element's bounds. The semi-Lagrangian method is hence local and parallel and
does not have the grid complexity, and parallelization challenges of the commonly used Lagrangian particle solvers in particle-mesh methods for solution of Eulerian-Lagrangian formulations. Formal proof is presented that the semi-Lagrangian algorithm evolves the solution according to the Eulerian Fokker-Planck equation. 
Numerical tests in one and two dimensions for drift-diffusion problems show that the method converges exponentially for constant and non-constant advection and diffusion velocities. 
\end{abstract}


\section{Introduction}
\label{sec:intro}
Chaotic dynamics govern the behavior
of a range of physics, such as  turbulent flows, molecular
dynamics, and plasmas. The  mathematical modeling of such stochastic physics requires a
formulation based on a multi-dimensional probability
density function (PDF). 
Molecular diffusion processes, for example,  are well-known
to be described by the Maxwellian PDF in phase space. 
In turbulence modeling the Fokker-Planck (FP) equations
 govern the  PDF  of sub-grid
velocity fluctuations and correlations thereof \cite{Givi99}.
Similarly, in chemically reacting turbulent flow
the FP model governs the probability density function
that is dependent on the number of species involved in the chemical reaction \cite{Givi99,Haworth2010}. 
Yet another example is the Vlasov model that describes the motion of charged
particles in phase space (See for example \cite{Birdsall}).

The dimensionality of the PDF in stochastic models is usually high.
In a chemical reaction, for example,  the number
of species, i.e. the dimension $d$ of the PDF, can easily be  on the  order of $100$. If we use $N$ degrees of freedom to approximate the stochastic partial differential equations that governs the PDF, then
the degrees of freedom required for solution is on the order of $N^d$. This
can easily yield  problem sizes  beyond the limitations of modern day computational infrastructure, even for simple problems with a relatively low  number of spatial dimensions.

In order to overcome this so-called "curse of dimensionality",  the  equations
that govern the model are usually not solved directly. 
Rather, a Monte-Carlo approach is used that provides samples from which the PDF
can be constructed. 
The Fokker-Planck equation for chemical species, for example, is commonly solved using  
an equivalent model based on a stochastic differential equation (SDE) \cite{Pope98}. 
In this approach, fictitious Monte Carlo(MC) particle tracers that
carry the species'  information are advected in physical space according to the SDE. 
At any time, the spatially dependent PDF is recovered using averaging techniques on the MC realizations.

In a similar fashion, it is well known that the diffusion equation can be solved with Monte-Carlo techniques
based on random walk models and stochastic Wiener processes. 
In grid based random walk (RW) methods \cite{Ames}, 
fictitious particles that represent the concentration field are seeded at equidistant grid points.
The grid points are spaced by a distance of $\sqrt{2D\Delta t}$, where $D$ is the diffusion coefficient and
$\Delta t$, a time increment. The random walking particles jump 
 to a neighboring grid point with equal probabilities. For
diffusion processes, this method can be proven
equivalent to the second-order central finite difference  approximation of the
second-order differential terms in diffusion equations  \cite{Ames}.

Strong RW methods fall into the broader class of particle methods 
\cite{TOMPSON98} that do not depend on an underlying mesh.
In the strong RW method the tracers are randomly initialized. They
move a distance of $\sqrt{2D\Delta t}$ over a time
$\Delta t$ according to a Wiener process, 
$dW_t$, that is defined by an independent random number selected from a normal distribution with a mean of zero. At any given time 
and point in space, a PDF can be determined from the tracers
through binning and/or distribution of the tracer's
influence using distribution functions.
Because the Monte-Carlo method is well-known to convergence according
to the inverse of the square root of the number
samples, a large number of samples is required.
To achieve an error on the order of $10^{-3}$, for example,
one million samples are required at each point in space. In practical simulations reported in literature, the number of samples is usually much smaller, and the sampling error is is of engineering accuracy within a few percent.

To reduce the computational cost and improve accuracy, Ref. \cite{Vamos03} proposed the so-called "global random walk"(GRW), which is a modification of the weak RW method. In GRW a share of the tracer particles are not moved  and the remaining share is scattered to the neighboring nodes according to a Bernoulli distribution. This reduces the number of required Monte-Carlo realizations since the particles are distributed according to a single random number. The GRW method generalizes to a finite difference method for diffusion processes and is generally limited to low-order accuracy in space.

In many physics models the stochastic Lagrangian  model couples to a system  of Eulerian partial differential equations that governs the field dynamics for the particle tracer.
In turbulence modeling, for example,
the stochastic tracer that models  the PDF of subgrid turbulence stresses is  coupled
to an averaged or filtered flow  model, i.e.
the filtered or averaged Navier-Stokes equations.
In plasmas, the Maxwell equations govern the electric and magnetic fields that
force the stochastic motion of charged particles and vice-versa.

High-order accurate schemes like discontinuous spectral element methods (DSEM) \cite{Kopriva, KoprivaJacobs}
are a particularly good choice to solve these time-dependent Eulerian
equations. 
Because of their low dispersion and diffusion errors DSEMs 
are generally better  at propagating  waves over longer distances
and they capture small scales with fewer degrees
of freedom as compared to low-order methods. 
Moreover, DSEM  approximates the governing Eulerian equations on
unstructured grids of quadrilateral or hexahedral elements
which allows for the simulation of
complex geometry.  Since the method is local, i.e. no overlap between
elements, DSEM is highly parallel.
DSEM Navier-Stokes solvers have been shown extensively to
obtain high accuracy and convergence using unstructured grids on complex geometries
 \cite{HW08, Kopriva09}.
Moreover, both in theory \cite{HW08}
and in testing through benchmarks (e.g. \cite{Kopriva,JKM04,HOworkshop}),
DSEMs have been shown to have superior computational efficiency and parallelism
for computation of smooth flows as compared to more traditional discretization methods.

Because of the dynamic nature of the tracers, 
the consistent high-order coupling of (Monte-Carlo)  tracer particles  to the DSEM
framework is challenging and computationally expensive. Several studies
report on the coupling of the stochastic tracers to a DSEM field solver.
In Refs. \cite{KoprivaJacobs,JH06,JH09,JD09,SJD14,SJ16,SS11}, 
consistent interpolation methods were developed. Several high-order distribution functions
were proposed to distribute the particle influence on to the Eulerian grid.
In Ref. \cite{Komperda20} linear distributions functions are used that are local to an element. 
Refs. \cite{SAMMAK2016158,Givi17} couples an ensemble average solution determined on hexahedral domains to the unstructured Eulerian DSEM solver. 
In all of these approaches, either accuracy and/or the locality of the method is compromised, which
is detrimental to the computational efficient solution of the model.

As an alternative to Lagrangian particles tracers,
we introduced an explicit high order semi-Lagrangian (SL) method
for the solution of deterministic transport equations in
Ref. \cite{NatarajanJacobs20}. The SL method solves the Lagrangian form of the transport equations
and can be used instead of the particle solver in Eulerian-Lagrangian formulations.
By seeding particles on the DSEM solver's quadrature nodes within a spectral element,
the connection between the particle and field solver is direct and consistently high-order accurate.
Particles are integrated one time step forward along their characteristic path. 
The time step is restricted such that particles do not
cross the element boundaries.
The advected particle solution is remapped to 
the collocation nodes using a least-squares
fit with boundary and mass conservation constraints. The  SL method thus remains local and does not require additional attention for parallel computing other than at the elements interfaces when the advected solution is patched at interfaces. 

In this paper, we adopt and test the semi-Lagrangian method for the Monte-Carlo solution
of probability density function equations and diffusion equations with a  stochastic differential equation.
Monte-Carlo tracers are tracked  stochastically with the semi-Lagrangian method developed in \cite{NatarajanJacobs20} and are sampled  at the Gauss quadrature points.
There is hence no need for a binning or a distribution method.
Because the method
is local, the SL approach ensures high parallel efficiency and
high order accurate boundary condition implementation that has
eluded and plagued SDE methods coupled with high-order field solvers thus far. The resulting approach shares some similarities with the Eulerian Monte Carlo method \cite{Valino98,Sabelnikov05}, but also has distinct advantages over that approach. 

The paper is organized as follows. First, generic stochastic differential
equations and their equivalent Eulerian forms are discussed. Next,
a staggered-grid discontinuous spectral element method for the solution of an Eulerian field is briefly summarized. Before introducing the semi-Lagrangian algorithm  for solution of the
stochastic differential equation, we review for reference some common random walk methods. 
Tests are conducted for one dimensional,  constant and non-constant diffusion
problems as well as for stochastic problems. Finally, the SL solver is coupled
with a Navier-Stokes solver for the solution of a species equation in a temporally
developing shear layer.
Conclusions and future steps are reserved for the final section.
\section{Governing equations}
\label{sec:gov_eqns}
We consider the canonical stochastic differential equation in the It\^{o} sense for transport of MC particles in the physical space, $\mathbf{X}_t$:
\begin{equation} \label{eq:canonical_sde}
d \mathbf{X_t} = \mathbf{u}(\mathbf{X}_t,t) dt + \sigma(\mathbf{X}_t,t) d\mathbf{W}_t.  
\end{equation}
This SDE is central to many models for a range of problem as described in the introduction, including the filtered mass density function
model for the modeling species transport in subgrid turbulent scales, which is the broader focus of our research \cite{Givi99}.  It is usually complimented by
a transport equation for a variable in compositional space
that we do not consider here.
In (\ref{eq:canonical_sde}),  the drift velocity, $\mathbf{u}$, interpolates from a coupled field solver.  For the problems we consider, these are usually the Navier-Stokes equations. In most of the test cases below, we will assume a prescribed $\mathbf{u}$ field.
 The diffusion is driven by a Wiener process $W_t$ with diffusion coefficient, $\sigma$.

 
The probability density, $\Pdf(\mathbf{x},t)$
can be recovered from  Monte-Carlo (MC) realizations of the Lagrangian stochastic differential equation at any given point in space and time. This MC solution  is well-known to be equivalent
to solving the Eulerian, Fokker-Planck equation for $\Pdf(\mathbf{x},t)$ (e.g. \cite{Pope98}) given by
\begin{equation} \label{eq:fokkerplanck}
    {\partial{\Pdf(\mathbf{x},t)} \over \partial t} +  {\partial{\left(u_j(\mathbf{x},t){\Pdf(\mathbf{x},t)}\right)}\over \partial x_j} 
    = {\partial^2 \left({D(\mathbf{x},t){\Pdf(\mathbf{x},t)}}\right) \over {\partial x_j \partial x_j}},
\end{equation}
with $D = \sigma^2/2$.

For a constant diffusion coefficient $D(\mathbf{x},t)=D_c=\rm{constant}$, we can rewrite 
(\ref{eq:fokkerplanck}) as
\begin{equation} \label{eq:conv-diffusionPrecursor}
    {\partial{\phi(\mathbf{x},t)} \over \partial t} +  u_j(\mathbf{x},t){\partial{{\phi(\mathbf{x},t)}}\over \partial x_j} 
    = D_c {\partial^2 {{\phi(\mathbf{x},t)}} \over {\partial x_j \partial x_j}} - 
    \phi(\mathbf{x},t){\partial{{u_j(\mathbf{x},t)}}\over \partial x_j},
\end{equation}
where we have used the notation $\phi(x,t)$=$\Pdf(x,t)$ to be consistent with the deterministic formulation and method as described \cite{NatarajanJacobs20}.
For a conservative medium with a divergence
free velocity field, the equation further reduces to the generic convection diffusion equation
\begin{equation} \label{eq:conv-diffusion}
    {\partial{\phi(\mathbf{x},t)} \over \partial t} +  u_j(\mathbf{x},t){\partial{{\phi(\mathbf{x},t)}}\over \partial x_j} 
    = D_c {\partial^2 {{\phi(\mathbf{x},t)}} \over {\partial x_j \partial x_j}}.
\end{equation}
We use the many known analytical solutions
for this equation to assess the accuracy of the semi-Lagrangian method for solution of the SDE below.

The new procedure introduced in this work is capable of solving a wide class of Fokker-Planck problems, including (\ref{eq:fokkerplanck}). Specifically, we solve for a random field, $\phi^* (\mathbf{x},t)$, whose PDF is denoted by $\Pdf_\phi (\psi; \mathbf{x},t)$ (where $\psi$ is the sample space variable of $\phi^*$). The semi-Lagrangian scheme then leads to the following Fokker-Planck equation for $\Pdf_\phi (\psi; \mathbf{x},t)$: 

\begin{equation} \label{eq:fokkerplanckPhi}
\frac{\partial \Pdf_\phi}{\partial t} + u_j \frac{\partial \Pdf_\phi}{\partial x_j} = \frac{\partial}{\partial x_j}\left( D \frac{\partial \Pdf_\phi}{\partial x_j} \right) - \frac{\partial}{\partial \psi} \left[ S \Pdf_\phi \right],
\end{equation}

\noindent where $S(\psi;\mathbf{x},t)$ is a deterministic source term defined over the sample space. Note that while (\ref{eq:fokkerplanckPhi}) is given for a one-dimensional random variable $\phi^*$, the extension to multi-dimensional ${\phi}^*$ is trivial. In the present work we set $S(\psi;\mathbf{x},t)\equiv -\psi \frac{\partial u_j}{\partial x_j}$ for the purpose of recovering (\ref{eq:conv-diffusionPrecursor}) for the case when $D=D_c=\rm{constant}$.

Setting $\phi(\mathbf{x},t) = \int \psi \Pdf_\phi(\psi; \mathbf{x},t) d\psi$ to be the mean of $\phi^*(\mathbf{x},t)$ at a specific point $(\mathbf{x},t)$, it can be easily shown that taking the first moment, $\int \psi \cdots d\psi$, of each term in (\ref{eq:fokkerplanckPhi}) leads to (\ref{eq:conv-diffusionPrecursor}). This allows us to compare our method with methods such as global, strong and weak RW, whose Fokker-Planck equation is (\ref{eq:fokkerplanck}).

While some Monte Carlo solvers aim to find a solution to (\ref{eq:fokkerplanck}), in many applications (\ref{eq:fokkerplanckPhi}) is just as useful as a starting point, and there is no need to go through (\ref{eq:fokkerplanck}). As an example, in the large eddy simulation/filtered mass density (LES/FMDF) method of Jaberi et al.\cite{Givi99}, in the limit as the filter size goes down to $0$, the FMDF transport equation (eq.29 of \cite{Givi99}) for a one-dimensional compositional variable becomes equivalent to 

\begin{equation} \label{eq:FMDFgoal}
\frac{\partial \left( \left\langle \rho \right\rangle \widetilde{\Pdf_\phi} \right)}{\partial t} + \frac{\partial}{\partial x_j} \left( \left\langle \rho \right\rangle \widetilde{u}_j \widetilde{\Pdf_\phi} \right) = \frac{\partial}{\partial x_j}\left( \Gamma \frac{\partial \widetilde{\Pdf_\phi}}{\partial x_j}\right) - \frac{\partial}{\partial \psi} \left[ \left\langle \rho \right\rangle S \widetilde{\Pdf_\phi} \right],
\end{equation}

\noindent where $\widetilde{\Pdf_\phi}$ is the Favre (i.e., density-weighted) PDF of $\phi^*(\mathbf{x},t)$, $\left\langle \rho \right\rangle$ and $\widetilde{u}_j$ are the mean density and Favre-averaged velocity, $\Gamma$ is the combination of turbulent and molecular diffusivity, and the source term $S$ combines the effects of the mixing model and chemical reaction. In Jaberi et al., the authors perform a Monte Carlo solution of (\ref{eq:FMDFgoal}) using fully Lagrangian particles which evolve by an SDE. The drift term of this SDE's spatial component contains a gradient of the diffusivity so as to make the Fokker-Planck equation for the particle system (essentially the multi-dimensional, anisotropic version of (\ref{eq:fokkerplanck})) equivalent to (\ref{eq:FMDFgoal}). Alternatively, (\ref{eq:FMDFgoal}) can be recovered from (\ref{eq:fokkerplanckPhi}) by setting $u_j = \widetilde{u}_j - \frac{1}{\left\langle \rho \right\rangle ^2} \frac{\partial \left\langle \rho \right\rangle}{\partial x_j}\Gamma$ and setting $D=\frac{\Gamma}{\left\langle \rho \right\rangle}$. With these definitions of $u_j$ and $D$, (\ref{eq:fokkerplanckPhi}) can be multiplied through by $\left\langle \rho \right\rangle$ and (\ref{eq:FMDFgoal}) follows, provided the density consistency condition,

\begin{equation} \label{eq:densityCons}
\frac{\partial \left\langle \rho \right\rangle}{\partial t} + \frac{\partial}{\partial x_j}\left( \left\langle \rho \right\rangle \widetilde{u}_j  \right),
\end{equation}

\noindent is satisfied, meaning that the definition of $\rho({\psi})$ must be such that the mean density satisfies the averaged continuity equation. The need to satisfy density consistency does not make the present method any more cumbersome than Lagrangian particle methods, which have their own density consistency conditions \cite{Muradoglu2001,Popov2014} that they must satisfy.

Therefore, while (\ref{eq:fokkerplanck}) yields good test cases of the semi-Lagrangian method, the Fokker-Planck equation (\ref{eq:fokkerplanckPhi}), which the semi-Lagrangian method solves naturally is just as useful for modeling probabilistic systems.
 
\if 
We consider the canonical stochastic differential equation in the It\^{o} sense for transport of MC particles in the physical space, $\mathbf{X}_t$:
\begin{equation} \label{eq:canonical_sde}
d \mathbf{X_t} = \mathbf{u}(\mathbf{X}_t,t) dt + \sigma(\mathbf{X}_t,t) dW_t.  
\end{equation}
This SDE is used in the models of a range of problem as described in the introduction, including the filtered mass density function
model for the modeling species transport in subgrid turbulent scales, which is the broader focus of our research \cite{Givi99}. 
In (\ref{eq:canonical_sde}),  the drift velocity, $\mathbf{u}$, interpolates from a coupled field solver.  For the problems we consider, these are usually the Navier-Stokes equations. In most of the test cases below, we will assume a prescribed $\mathbf{u}$ field.
 The diffusion is driven by a Wiener process $W_t$ with diffusion coefficient, $\sigma$. 
Along the stochastic tracer, a generic transport property, $\phi$, can change in the compositional space as follows, 
\begin{equation} \label{eq:canonical_phi}
{d {\phi(X(t),t)} \over dt} = -\phi(X(t),t) {\partial \mathbf{u}(x,t) \over \partial x} +(\phi(X(t),t)-\langle \phi \rangle)+ S(\phi(X(t),t)).
\end{equation}
The first term on the right hand side of this equation stems from the advection of $\phi$ along the characteristic path according to the convection equation in conservation form, i.e in deterministic form following \cite{NatarajanJacobs20}
\begin{eqnarray}
{\partial \phi \over \partial t} + 
{\partial (\phi u) \over \partial x} & =&  \\
{\partial \phi \over \partial t} +u{\partial \phi \over \partial x} + 
\phi {\partial u \over \partial x} &= & \\
{D \phi \over Dt}  + \phi {\partial u \over \partial x} &=& 0,
\end{eqnarray}
and thus along the Lagrangian path it follows that
\begin{equation}
{D \phi \over Dt}  =- \phi {\partial u \over \partial x} 
\end{equation}
with $D \over Dt$ the total time derivative. For most conservative velocity the divergence of the velocity is zero and thus the right hand side is zero.
The second term relates to diffusion of $\phi$ in compositional space. The last term is a generation term that we assume is zero in this write-up, $S$=0.

 
The probability density, $\Pdf(\psi;x,t)$, can be recovered from  Monte-Carlo (MC) realizations of the Lagrangian stochastic differential equation at any given point in space and time. This MC solution  is well-known to be equivalent
to solving the Eulerian, Fokker-Planck equation for $\Pdf(\psi;x,t)$ (e.g. \cite{Pope98}) given by
\begin{equation} \label{eq:fokkerplanck}
    {\partial{\Pdf(\psi;x,t)} \over \partial t} +  {\partial{\left(u(x,t){\Pdf(\psi;x,t)}\right)}\over \partial x} 
    = {\partial^2 \left({D(x,t){\Pdf(\psi;x,t)}}\right) \over {\partial x^2}} + {\partial \left(\psi - \langle \phi \rangle \Pdf \right)\over  \partial \psi },
\end{equation}
with $D = \sigma^2/2$.

  The average value of the transport variable, $\langle\phi(x,t)\rangle$ at any given location, $x$, can be obtained by taking the first moment of the PDF as follows:
  \begin{equation}
  \langle{\phi(x,t)}\rangle=\int{\psi \Pdf(\psi;x,t)}d\psi
  \end{equation}
  Similarly we can take the first moment of (\ref{eq:fokkerplanck}) and obtain
  \begin{equation} \label{eq:canonical_pde}
     \frac{\partial \langle \phi(x,t) \rangle}{\partial t} +  u(x,t) \frac{\partial  \langle{\phi(x,t)}\rangle}{\partial x} = D_c \frac{\partial^2 \langle\phi(x,t)\rangle}{\partial {x}^2}
\end{equation}
under the condition that $S$=0, $u(x,t)=\langle u(x,t)|\psi=\phi(x,t)\rangle$  
for a conservative medium ($\partial{u} /\partial{x}$=0) and a constant
diffusion coefficient $D(x,t)=D_c=\rm{constant}$. This is the generic convection diffusion equation. We use the many known analytical solutions
for this equation to assess the accuracy of the semi-Lagrangian method below.

\fi


\section{Discontinuous Spectral Element Method}
\label{sec:dsem}
Following \cite{NatarajanJacobs20}, the  semi-Lagrangian (SL) method is consistently
coupled to the staggered grid discontinuous spectral element method (DSEM) 
as first introduced by Kopriva \cite{Kopriva}.
In this version of DSEM the  solution variable is collocated
at Gauss quadrature nodes and the fluxes on Lobatto quadrature nodes.
The collocation at Gauss quadrature nodes are specifically beneficial
for the simple and consistent coupling between the SL method and DSEM because it
it leads to preservation of the high-order, local nature of DSEM 
as we showed in \cite{NatarajanJacobs20}.  For completeness, we briefly
summarize essential aspects of the staggered grid DSEM method again.  For a detailed
description, we refer to \cite{Kopriva09,JKM04,NatarajanJacobs20}.

In DSEM, the physical domain $\Omega$ is divided
into $K$ non-overlapping elements, $\Omega = \cup_{k=1}^K \Omega_k $. 
In the context of DSEM, elements are often referred to as subdomains, a nomenclature
that we follow in this paper. Each physical
subdomain is then mapped onto a unit computational cube using iso-parametric
transformation \cite{KoprivaJacobs}. The governing Eulerian equation is given by, 
\begin{equation} 
\frac{\partial \tilde{\mathbf{Q}}}{\partial t} + \tilde{\nabla}\cdot\tilde{F} =0,
\label{eq:3.1}
\end{equation}
where, $\tilde{\mathbf{Q}}= |\overline{\overline{J}}| \mathbf{Q}$, $\tilde{\nabla}\cdot\tilde{F} = \frac{\partial \tilde{f}}{\partial \xi} + \frac{\partial \tilde{g}}{\partial \eta} + \frac{\partial\tilde{h}}{\partial \zeta}$. $|\overline{\overline{J}}|$ is the determinant of the transformation from the physical to the computational domain. 

The solution and flux collocation points are chosen according to Chebyshev Gauss and Lobatto quadrature points,
which along tensorial grid lines,  $0\leq \xi \leq 1$, are given by,
\begin{equation} \label{eq:gaussquad}
\xi_{i+1/2} = \frac{1}{2}\left[ 1- \cos\left(\frac{i+1/2}{N+1}\right) \pi \right] \qquad i=0,1,...,N-1,
\end{equation}
and
\begin{equation} 
\xi_i = \frac{1}{2}\left[1-\cos\left(\frac{i \pi}{N}\right)\right] \qquad i=0,1,...,N,
\label{eq:gausslobatto}
\end{equation} 
respectively.
Here, we have used the integer subscript, $i$, to identify Lobatto
points and $i+1/2$ to identify Gauss points that are located
in between two Lobatto points $i$ and $i+1$.
In three dimensions, the solution interpolant $\tilde{\mathbf{Q}}$ is then
\begin{equation} 
\tilde{\mathbf{Q}}(\xi, \eta, \zeta) = \sum \limits_{i=0}^{N-1} \sum \limits_{j=0}^{N-1} \sum \limits_{k=0}^{N-1} {\tilde{\mathbf{Q}}}_{i+1/2,j+1/2,k+1/2} h_{i+1/2}(\xi) h_{j+1/2}(\eta) h_{k+1/2}(\zeta),
\label{eq:3.4}
\end{equation}
where $h_{i+1/2}(\xi)$ is the Lagrange interpolation polynomial of degree N-1 defined on the Gauss 
quadrature points $\xi_{m+1/2}$ 
and
\begin{equation} 
h_{i+1/2}(\xi) = \prod_{\substack{m=0 \\ m \neq p}}^{N-1} \frac{\xi - \xi_{m+1/2}}{\xi_{i+1/2} - \xi_{m+1/2}}, \qquad i=0,1,...,N-1,
\label{eq:3.5}
\end{equation}
is the Lagrangian polynomial of degree $N$-1.
The fluxes, $\tilde{F}$, are collocated  similarly on the Lobatto points. Through interpolation 
between the Gauss grid and the Lobatto grid, the fluxes can be determined as a function
of the solution, $\tilde{Q}$.  Through an approximated Riemann solver,
an interface flux is determined from interface solutions on neighbouring subdomains.
The derivatives of the  fluxes, $\tilde{\nabla}\cdot\tilde{F}$, are  determined at the Gauss points.
Then, it remains to update the Gauss solution  in time. We typically use an
explicit integrator such as a  standard fourth order explicit Runge-Kutta time stepping method.

\section{Stochastic Random Walk based Methods}
\label{sec:prev_methods}
Before we present the semi-Lagrangian methods based on DSEM, we review some
of the most common random walk methods in one-dimension which we will use for comparison and reference to the SL-DSEM. 

\subsection{Strong random walk method}

In the strong random walk method, 
the spatial location $x_j^{p}$ of $N_p$ Monte Carlo tracers are advected 
according to the SDE in (\ref{eq:canonical_sde}) using the first order Euler-Maruyama method \cite{Kloeden}
as follows:
\begin{equation} \label{eq:lagrangian_particle}
 x_j^{p}(t+\Delta t) = x_j^{p}(t) + \Delta t \mathbf{u}(x_j^{p}(t))  + \sqrt{2D} dW_t \ \ \ \ j=1...N_p.  
\end{equation}
Here, $\Delta t$ is the time step. The particle's solution
in compositional space, $\phi_j^{p}$, is advected along its characteristic path.
The particles are randomly
seeded and traced within a computational domain defined on the interval $[x_a, x_b]$.

By sampling within bins (or elements)  the probability
density function, $\Pdf(x,t)$ can be constructed. 
We use $N_b$ equidistant bins between $x_i$ and $x_{i+1}$ with the center location of each bin given by
with
\begin{equation}
x_{i+1/2} = i \Delta x + x_a+ {\Delta x \over 2}\ \ \ \ i=0....N_b-1
\end{equation}
and $\Delta x=(x_b-x_a)/(N_b)$.

The analytical average,  $\langle\phi\rangle(x_{i+1/2})$ within a bin with center location
$x_{i+1/2}$, is determined using the first moment of $\Pdf(x,t)$ with respect to $\phi$ as
\begin{equation}
\langle{\phi}\rangle_{i+1/2} = \int_{x_i}^{x_{i+1}} \phi(x)  \Pdf(x_{i+1/2},t) dx
\end{equation}
This is equivalent to ensemble averaging of  the MC realizations
within a bin.

 
 \subsection{Weak random walk methods}
 The weak random walk algorithm is grid based (e.g. \cite{Ames}).
Starting again from a computational domain on the interval $[x_a, x_b]$, we
define an equidistant grid with $N_b$ equidistant elements using the nodes,

\begin{equation} \label{eq:1d_grid}
x_i = i \Delta x + x_a\ \ \ \ i=0....N_b,
\end{equation}
where $\Delta x=(x_b-x_a)/(N_b)$ is the grid spacing.

Total number of particles, $N_p$ is
uniformly distributed over the nodes. The
number density weighted solution at a given node is initialized as
\begin{equation} \label{eq:part_distribution}
    m^{0}_i = {1 \over 2} \phi^{0}_i \Delta x N_p \ \ \ \ i=1....N.
\end{equation}
with $\phi^{0}_i$ the initial condition.
Each particle at a given node moves to a neighboring node with equal probability
from time $t^n$ to $t^{n+1}$.
The time step size is $\Delta t=t^{n+1}-t^{n}$, and it must be
related to the grid spacing $\Delta x$ according to,
\begin{equation}
    \Delta x = \sqrt{2D \Delta t}.
\end{equation}
to consistently capture the diffusion in (\ref{eq:conv-diffusion}). 
The updated distribution of particles at time step $n+1$ can be determined as,
\begin{equation}
    m^{n+1}_i = {1 \over 2}(m^n_{i-1} +m^n_{i+1}). 
\end{equation}
At the new step, $\phi_i^{n+1}$, is recovered  as follows
\begin{equation} \label{eq:rw_redistribute}
    \phi^{n+1}_i = {{2 m^{n+1}_i} \over {\Delta x N_p}}                    \ \ \ \ i=1....N.
\end{equation}
 
 \subsection{Global random walk method (GRW)}
 
The GRW method introduced in \cite{Vamos03} is similar to the random walk method. The GRW method does not move
individual particles with equal probability, however. Instead it moves particles in large groups
according to a prescribed probability density function to reduce the number of samples
and thus reduce computational cost.
 The domain and the initial particle distribution are the same as for the random walk method given in
  (\ref{eq:1d_grid}) and (\ref{eq:part_distribution}),
  respectively. 
 Let $\delta m^n(j,i)$ denote the density weighted solution for a group of particles at time $t^n$ moving from node $x_j$ to $x_i$.
 For a given time step only a fraction $r$ of the number of particles move to the neighboring nodes, the rest of them determined by
 \begin{equation} \label{eq:grw_rhs1}
     \delta m^n(i,i) = (1-r) m^n_i       \ \ \ \ i=1....N.
 \end{equation}
 remain at the same node.
   The parameter $r$ connects $\Delta x$ and $\Delta t$ as follows
 \begin{equation}
     r= {{2D \Delta t} \over {(\Delta x)^2}}
 \end{equation}
 to ensure a consistent solution of (\ref{eq:conv-diffusion}). 
 Assuming the particles are moved only to the nearest neighboring node, the distribution of particles at $t^{n+1}$ for a given node $x_i$ is determined as 
 \begin{equation} \label{eq:grw_distribution}
     m^{n+1}_i = \delta m^n(i,i) + \delta  m^n(i+1,i) + \delta m^n(i-1,i) \ \ \ \ i=1....N.
 \end{equation}
The second and third term on the right hand side of this equation 
represent contributions from the the neighbouring nodes.
The neighbouring groups move according to a Bernoulli
distribution given by $b_m(\alpha)= 2^{-m} C_m^\alpha$.
This distribution is sampled by a random number generator to provide $\alpha$  so that
 \begin{equation} \label{eq:grw_rhs2}
     \delta m^n(i,i+1) =\alpha,     \ \ \ i=1....N,
 \end{equation}
 and
 \begin{equation} \label{eq:grw_rhs3}
     \delta m^n(i,i-1) = m^n_i - \delta m^n(i,i) -\alpha,     \ \ \ i=1....N,
 \end{equation}
 Using $m_i^{n+1}$ in (\ref{eq:grw_distribution}),  $\phi_i^{n+1}$ is again recovered using (\ref{eq:rw_redistribute}).


\subsection{Eulerian Monte Carlo Method (EMC)}

The EMC method, first developed by Valiño \cite{Valino98}, and refined by Sabel'nikov and Soulard \cite{Sabelnikov05} is another alternative for solving Fokker-Planck equations. In contrast to the abovementioned particle methods, EMC solves for the PDF of $\phi^*$ by tracking a set of fields $\Phi^{(s)}$, for $s=1,...,N_f$, which are defined on the entire domain. The fields $\Phi^{(s)}$ are evolved by a stochastic partial differential equation (SPDE) such as the following:

\begin{equation} \label{eq:emc_spde}
d\Phi^{(s)} + u_j \frac{\partial \Phi^{(s)}}{\partial x_j} dt + \frac{\partial \Phi^{(s)}}{\partial x_j} \sigma dW_j^{(s)} - \frac{\partial}{\partial x_j}\left(  D \frac{\partial \Phi^{(s)}}{\partial x_j}  \right) dt = S(\Phi^{(s)};\mathbf{x},t)dt,
\end{equation}

\noindent where $D=\sigma^2 /2$ and $S(\phi; \mathbf{x},t)$ is a general source term. In the context of reactive flow simulations, this source term will be a combination of the reaction source term and the effect of molecular diffusion. We note that the Wiener increments $dW_j^{(s)}$ are spatially global, i.e., the same Wiener increment sample $dW_j^{(s)}$ is used for all points $\mathbf{x}$ in (\ref{eq:emc_spde}).

Following \cite{Sabelnikov05}, (\ref{eq:emc_spde}) leads to the Fokker-Planck equation

\begin{equation} \label{eq:emc_FP}
\frac{\partial \Pdf_\phi}{\partial t} + \frac{\partial}{\partial x_j}\left( u_j \Pdf_\phi \right) = \frac{\partial}{\partial x_j} \left(  D \frac{\partial \Pdf_\phi}{\partial x_j}  \right) - \frac{\partial}{\partial \psi} \left( S(\psi) \Pdf_\phi \right),
\end{equation}

\noindent where $\Pdf_\phi(\psi;\mathbf{x},t)$ is the PDF of $\Phi(\mathbf{x},t)$ at specified values of $\mathbf{x}$ and $t$, and $\psi$ is the sample space variable of $\Phi$. It is easily seen that (\ref{eq:emc_FP}) is equivalent to (\ref{eq:fokkerplanckPhi}).

Like  EMC,  DSEM-SL  determines its ensemble of smooth fields by applying the same Wiener increment to all points in a given sample field. 
As a result, in the limit of arbitrarily high spatial resolution, DSEM-SL solutions converge to the same SPDE solved by EMC. This is formally  proven in Appendix A. 

A downside to EMC is  the appearance
of the term  $\frac{\partial \Phi^{(s)}}{\partial x_j} \sigma dW_j^{(s)}$ of (\ref{eq:emc_spde}) in the formulation. The combinination of  a derivative approximation with the Wiener increment $dW_j^{(s)}$, 
is particularly challenging. 
DSEM-SL avoids this term and has other advantages that
will be discussed in the presentation of DSEM-SL in the next section.

Finally, EMC methods have so far been implemented only with low-order FV \cite{Valino98} and ENO \cite{Sabelnikov05} spatial discretizations, whereas DSEM-SL exhibits spectral spatial convergence.

\section{Semi-Lagrangian method for stochastic differential equation}
\label{sec:dsem_sl}
The semi-Lagrangian algorithm for simulation
of the stochastic differential equation is based
on the semi-Lagrangian method that we developed in \cite{NatarajanJacobs20} for the Monte-Carlo simulation
of deterministic Lagrangian transport equations in Eulerian-Lagrangian formulations.
To solve stochastic models using Monte-Carlo sampling from tracers
that behave according to the stochastic
differential equation, multiple polynomial
solutions are generated with the deterministic semi-Lagrangian method
 according to a Wiener process. Similar to the strong random walk method, each
polynomial realization represents
a Monte-Carlo sample and can be used to reconstruct
the density function at quadrature points. Below,
we discuss the semi-Lagrangian method in one-dimension and highlight
the implementation of stochastic components. The multi-dimensional algorithm can be formulated on a tensorial
grid as discussed in \cite{NatarajanJacobs20}. For brevity, we refer for details for the multi-dimensional algorithm to that article.

\subsection{Solution initialization}

To be consistent with the Eulerian DSEM solver that provides
the drift velocity, $u$,  at the tracer location in (\ref{eq:canonical_sde}), 
we initialize $N_p=N$ particles within a subdomain, $\Omega_k$, at an initial time $t^0$ at the $N$ Gauss quadrature points in (\ref{eq:gaussquad}). 
The drift velocity is directly available at these quadrature points
and hence does not require computational intensive interpolation
that is necessary for general Lagrangian particle methods.
A single sample, $s$, of the solution, $\phi^{s}$, at a given time $t^n$ is approximated by a Lagrange interpolant as follows,
\begin{equation} \label{eq:soln_init}
	{\phi^{s}}^n(\xi) = \sum \limits_{i=0}^{N-1}{{\phi}^s}^n({\xi^s}^n_{i+1/2}) h_{i+1/2}(\xi), \ \ s=1,..,N_s
\end{equation} 
where $N_s$ is the total number of Monte-Carlo samples used per grid point.
The Lagrange polynomials, $h_{i+1/2}(\xi)$ , of degree $N-1$ are defined on the Chebyshev Gauss points $\xi_{i+1/2}^s$ according to (\ref{eq:gaussquad})
\begin{equation} \label{eq:lagrange_poly}
h_{i+1/2}(\xi) = \prod_{\substack{i=0 \\ i \neq j}}^{N-1} \frac{\xi - {\xi^s}^n_{i+1/2}}{{\xi^s}^n_{j+1/2} - {\xi^s}^n_{i+1/2}}, \qquad j=0,1,...,N-1.
\end{equation}
 

\subsection{Forward time integration}
The particles and its associated sample polynomial solution, $\phi^{s}$, are advected in the physical space along its characteristic path according to the stochastic differential equation (\ref{eq:canonical_sde}).
To integrate the SDE in It\^{o} form, we use an explicit first-order Euler-Maruyama scheme\cite{Kloeden}, so that in local coordinates the time step will have the form
\begin{equation} \label{eq:advected_particles}
	 \xi^{s^\star}_{i+1/2} = {\xi^{s}}^n_{i+1/2}+\left(\Delta t \left.\left[ {u + D \frac{\partial^2 \xi}{\partial x ^2}\frac{ \partial x }{ \partial \xi}}\right]\right\vert_{{\xi^{s}}^n_{i+1/2}}   + \sqrt{2D} \Delta W_t^s \right) / \frac{\partial x}{\partial \xi} .
     \ \ i=0,1,...,N-1,
\end{equation}

\noindent and in physical coordinates its form will be

\begin{equation} \label{eq:def_dx}
  x^{s*}_{i+1/2} = x_{i+1/2} + u \left( x_{i+1/2}, t \right)\Delta t + \sqrt{2D\left( x_{i+1/2}, t \right)} \Delta W_t^s.
\end{equation}

Here, $\Delta W_t^s$ is the Wiener increment that is obtained from a random number generator according to a normal distribution with a mean of zero and a variance of $\Delta t$. As indicated by the notation $\Delta W_t^s$, the same sample of the Wiener increment is used for the advection of all spatial points which belong to the solution $\phi^s$. This is similar to what is done in Eulerian Monte Carlo, and in contrast to a scheme such as the Lagrangian particle method, in which each new point gets its own sample of the Wiener increment. The advantages of the present approach are twofold: firstly, it preserves the spatial smoothness of $\phi^s$, which is required for the correct convergence of the spatial discretization schemes. Secondly, it reduces computational effort, as much fewer calls to the random number generator are needed.

The above advection formulations both yield convergence to the same result and are, in fact, identical for non-curvilinear grids. Here we use the physical coordinate formulation, (\ref{eq:def_dx}), which is preferable because it avoids the need to compute the higher-order metric term in (\ref{eq:advected_particles}). We note that this avoidance of metrics is a significant advantage of the semi-Lagrangian method over Eulerian Monte Carlo, as it reduces computational effort and yields a procedure which is better behaved on singular or close to singular grids.


To obtain high-order accuracy for the time integration of the It\^{o} form SDE
one can consider Runge-Kutta methods as discussed in \cite{Kloeden}. 
The algorithms extends naturally from the Euler-Maruyama to high-order time-integrators as we have shown 
in\cite{NatarajanJacobs20}. For SDEs, however, these high-order time-integrators are increasingly complex with increasing order and a topic of ongoing research. We have not considered them in this work, but aim to report on this in future investigations.




While there is no formal stability criterion for the temporal update of the linear characteristic equation,  we  prevent an advected particle from leaving a subdomain by restricting the time step. 
This has two reasons.
Firstly, if the advected particle locations within a subdomain deviates only marginally
from the quadrature point locations,  then the Vandermonde interpolating matrix
can be expect to be reasonable well-conditioned ensuring that the remapping
which requires an inversion of this matrix is not singular.
Secondly, the nodes stay local to the element, which means that the method
is local and parallel and that connection of the interfaces can be performed in 
relatively simple manner using the interpolation method described below. 

Thus, the time step restriction is set by the following condition,
\begin{equation} \label{eq:1d_timestep}
  |u|_{max} \Delta t + \sqrt{2D\Delta t} \leq \Delta x_{min}
\end{equation}
where, $\Delta x_{min}$ is the minimum grid
spacing between two particles in the physical space and $|u|_{max}$ is the maximum advection speed. For a pure diffusion problem without drift, this reduces to,
\begin{equation} \label{eq:1d_difftimestep}
  \Delta t \leq {(\Delta x_{min})^2 \over 2D},
\end{equation}

\noindent which is equivalent to a Fourier number condition, but one that can be violated without loss of stability. In contrast to the present scheme, the Wiener increment appearing in the advection term in EMC schemes yield a  Fourier number stability condition . Thus, another advantage (albeit one which is not used here) of semi-Lagrangian schemes is the ability to take larger time steps. 

The solution after advection is denoted by $\phi^{s^\star}(\xi)$ and is given as,
\begin{equation}
	\phi^{s^\star}(\xi) = \sum \limits_{i=0}^{N-1}{\phi^\star}(\xi^{s^\star}_{i+1/2}) h^{s^\star}_{i+1/2}(\xi), 
\label{eq:phi_stardef}
\end{equation}
where $h^{s^\star}_{i+1/2}(\xi)$ are the Lagrange  polynomials of degree $N-1$ 
defined on the advected points $\xi_{i+1/2}^{s^\star}$,
\begin{equation} 
h^{s^\star}_{i}(\xi) = \prod_{\substack{i=0 \\ i \neq j}}^{N-1} \frac{\xi - \xi^{s^\star}_{i+1/2}}{\xi^{s^\star}_{j+1/2} - \xi^{s^\star}_{i+1/2}}, \qquad j=0,1,...,N-1.
\label{eq:h_star}
\end{equation}

In general the advected polynomial's nodal solution values, ${\phi^{s^\star}}(\xi_{i+1/2}^{s^\star})$ is obtained by integrating $\phi^{s}(\xi_{i+1/2}^s)$ in the compositional space according to (\ref{eq:fokkerplanckPhi}) and the advection of $\phi$ along the flow,
\begin{equation} \label{eq:advection_phi}
 		{\phi^{s\star}}(\xi^{\star}_{i+1/2}) = {\phi}^s_{i+1/2} + \Delta t \left( -{\phi}^s_{i+1/2} \left(\frac{\partial u}{\partial x}\right)_{i+1/2}\right).
 \end{equation}
 
 Here, $\left(\frac{\partial u}{\partial \xi}\right)_{i+1/2}$ is obtained from the DSEM field solver solver.
\subsection{Remapping}
In the final remapping stage of the algorithm, the advected polynomial is projected back onto the Gauss-Chebyshev quadrature nodes through interpolation as follows,
\begin{equation} \label{eq:phi_star}
	\widehat{\phi}^{s^{n+1}}(\xi_{j+1/2}^s)
	 = \sum \limits_{j=0}^{N-1}{\phi^{s^\star}}(\xi^{s^\star}_{j+1/2}) h^{s^\star}_{j+1/2}(\xi^s_{i+1/2}), 
\end{equation}
Here, we use the \textit{hat} symbol to denote the intermediate solution at $t^{n+1}$.
To account for connectivity between elements and boundary conditions, we constrain this intermediate solution following \cite{NatarajanJacobs20}.
Boundary conditions and interface constraints are applied using interpolation. We determine the boundary values using polynomial interpolation according to (\ref{eq:phi_star}),
\begin{eqnarray} \label{eq:remap_bnd}
	\hat{\phi}^{s^{n+1}}_b = \sum \limits_{j=0}^{N-1}{\phi^{s^\star}}(\xi^{\star}_{j+1/2}) h^\star_{j+1/2}(\xi_b)  \qquad b=1,2 
\end{eqnarray}
By upwinding, a unique interface value  is determined
from the interfaces values of two neighbouring subdomains. If $u \Delta t + \sqrt{2D}dW_t$ is positive at the interface, then we use the information from the left element.  
\begin{eqnarray} \label{eq:bnd_update}
	\hat{\phi}^{s^{n+1}}_{b\star} =  f \left(\hat{\phi}^{s^{n+1}}_{b=1} \bigg\vert_{\Omega_k}, \hat{\phi}^{s^{n+1}}_{b=2} \bigg\vert_{\Omega_{k-1}} \right)
\end{eqnarray}
Boundary conditions are implemented in the same way as interface
condition by using a specified ghost solution at computational domains boundaries.

To project the interpolated polynomial, $\hat{\phi}^{s^{n+1}}$, combined with the boundary constraints onto the Gauss-Chebyshev quadrature we use a least-squares method to solve the overdetermined system of equations as described in \cite{NatarajanJacobs20}.

We note here that, for a multi-dimensional ${\boldsymbol{\phi}}^*$, the majority of the computational cost of the remapping stage does not scale up with increasing dimension of $\boldsymbol{\phi}^*$, since each component of $\boldsymbol{\phi}^*$ is defined on the same advected points. This is a significant advantage over EMC methods, for which the spatial discretization has to be applied to each component of the random field, and thus scales linearly with the random variable's dimension.


\subsection{Averaging}

To determine the mean of the polynomial solution, we ensemble average at the grid points only as follows:
\begin{equation} \label{eq:averaged}
   \langle\phi\rangle^{n+1}_{i+1/2} = {1 \over N_s} \sum \limits_{s=1}^{N_s} \phi^{s^{n+1}}_{i+1/2}  \qquad i=0,1,...,N-1.
\end{equation}
Because the averaging is performed on the quadrature nodes, this is equivalent to averaging the polynomial
on each element. We can also recover the PDF of the solution on each grid point by binning the samples. 

When using Dirichlet boundary conditions, the samples can develop high gradients at the boundaries. By re-seeding the samples from the averaged solution once every few time steps we can reduce the high gradient. In this paper we perform re-seeding after every 100 time steps. 
\begin{equation} \label{eq:re-seed}
       \phi^{s^{n+1}}_{i+1/2} = \langle\phi\rangle^{n+1}_{i+1/2}  \qquad i=0,1,...,N-1, \qquad s=1,2,...,N_s.
\end{equation}

\subsection{Consistency and Accuracy}

It can be proven that DSEM-SL is consistent, i.e. it solves the Fokker-Planck equation (\ref{eq:fokkerplanckPhi}) implicitly for the PDF of $\phi^*$. This proof and the derivation of the equivalent stochastic PDE is presented in Appendix A.

The accuracy of the method depends on three known approximation errors that include, 
(1)  the number of samples,  (2) the accuracy of the spatial approximation and (3) time integration accuracy.
The sampling error converges according to
the inverse of the square root of the number of samples. The spatial approximation is spectrally
accurate according to the high-order approximation in each element. We use a Euler-Maruyama time integration method, which is first order and hence the time integration accuracy is of $\mathcal{O}(\Delta t)$. In the test cases below, we fix the $\Delta t$ value and study the convergence of interpolation and the sampling errors. We confirm that the DSEM-SL converges
according to these expected error estimates.

\section{Numerical tests}
\label{sec:tests}
We assess the error and the behavior of the semi-Lagrangian method
for several one and two dimensional test cases.
Results are compared to the analytical solutions
and solutions obtained with classic random walk method and Eulerian Monte Carlo method as described above.

Accuracy is measured using the $L_2$ norm of the solution error which is calculated by summing up local $L_2$ error norms in each subdomain, $k$, as,
\begin{equation} \label{eq:l2_error}
\|e\|_{L^2} = \sum \limits_{k=1}^{K} \sqrt{\int_{\Omega_k} (\phi-\phi_{\text{exact}})^2 \mathcal{J} d\xi},
\end{equation}
where $\mathcal{J}$ is the Jacobian for the transformation from the physical space to the computational space.
We also asses conservation properties of the method are by inspecting the following global mass and energy norms, 
\begin{equation} \label{eq:mass_cons}
   \|M\|=\sum \limits_{k=1}^{K} \frac{\int_{\Omega_k} \phi d\xi}{\int_{\Omega_k}\phi_\text{exact}  d\xi},
\end{equation}
and
\begin{equation} \label{eq:energy_cons}
   \|E\| = \sum  \limits_{k=1}^{K} \frac{\int_{\Omega_k}(\phi)^2 d\xi}{\int_{\Omega_k}\phi^2_\text{exact} d\xi}.
\end{equation}
respectively.

\subsection{One dimensional constant diffusion: Sine function}
As a first test, we consider the diffusion of a sine wave  with the drift velocity $\mathbf{u}$ set to zero and the diffusion coefficient set to $D_c$=1 according to (\ref{eq:conv-diffusion}).
In the  domain   $x$=$[0,1]$ the initial condition is set to $\phi(x,0)$ =$\sin(2\pi x)+2$. Periodic boundary conditions are specified. In order to keep the time integration error low and to satisfy the stability criterion (\ref{eq:1d_difftimestep}), a time step of $\Delta t$= $10^{-5}$ is used. Simulations are carried out for 50 time steps for different number of element sizes, $H$, different polynomial orders, $P=N-1$ and different number of samples, $N_s$.

\begin{figure} 
\centering 
\mbox{
    \includegraphics[width=0.5\textwidth]{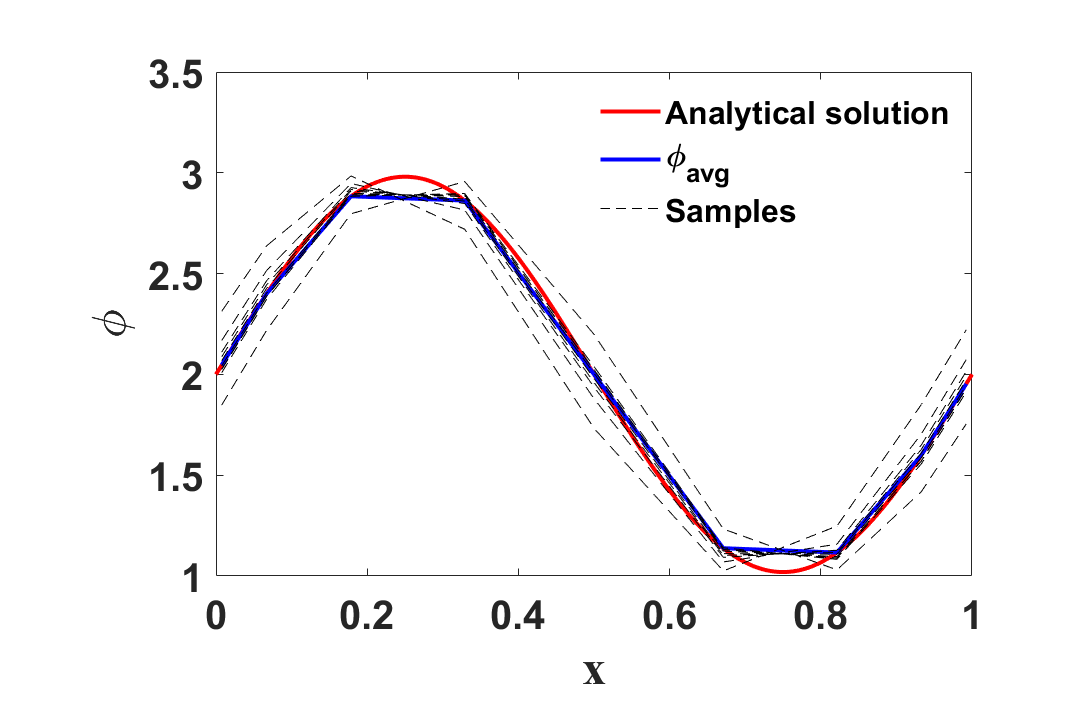}
    \includegraphics[width=0.5\textwidth]{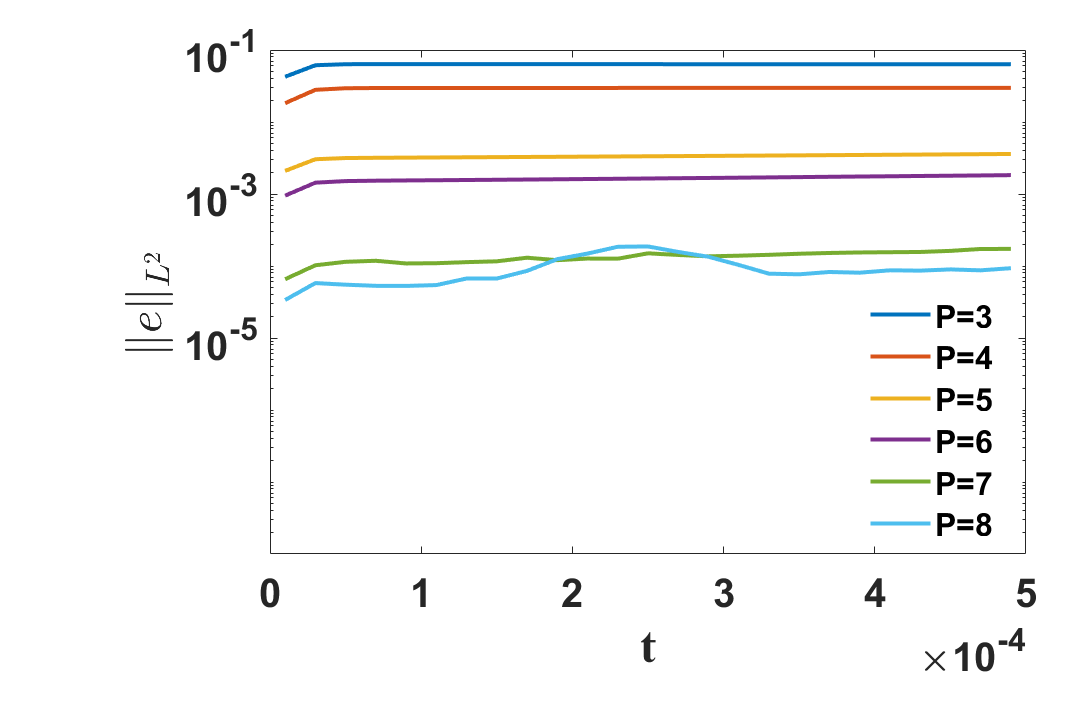} 
}
\mbox{
 \hspace{0.45cm}
 \makebox[0.45\textwidth]{(a)}
  \hspace{0.05\textwidth}
 \makebox[0.45\textwidth]{(b)}
 }
\centering
\mbox{
      \includegraphics[width=0.5\textwidth]{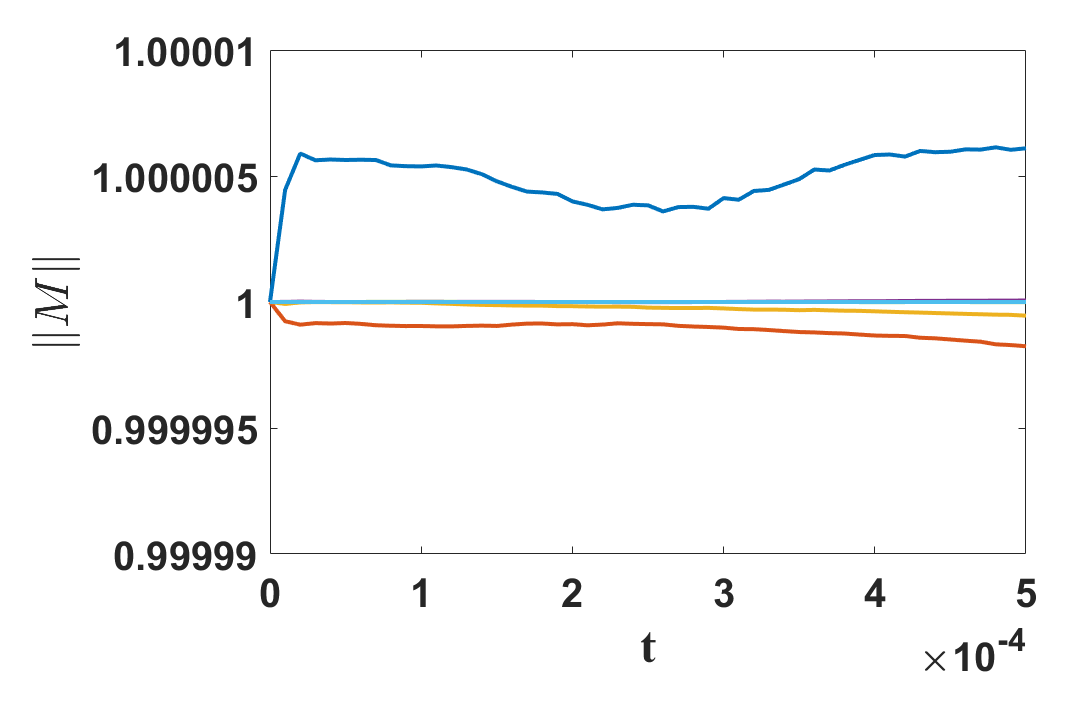}
     \includegraphics[width=0.5\textwidth]{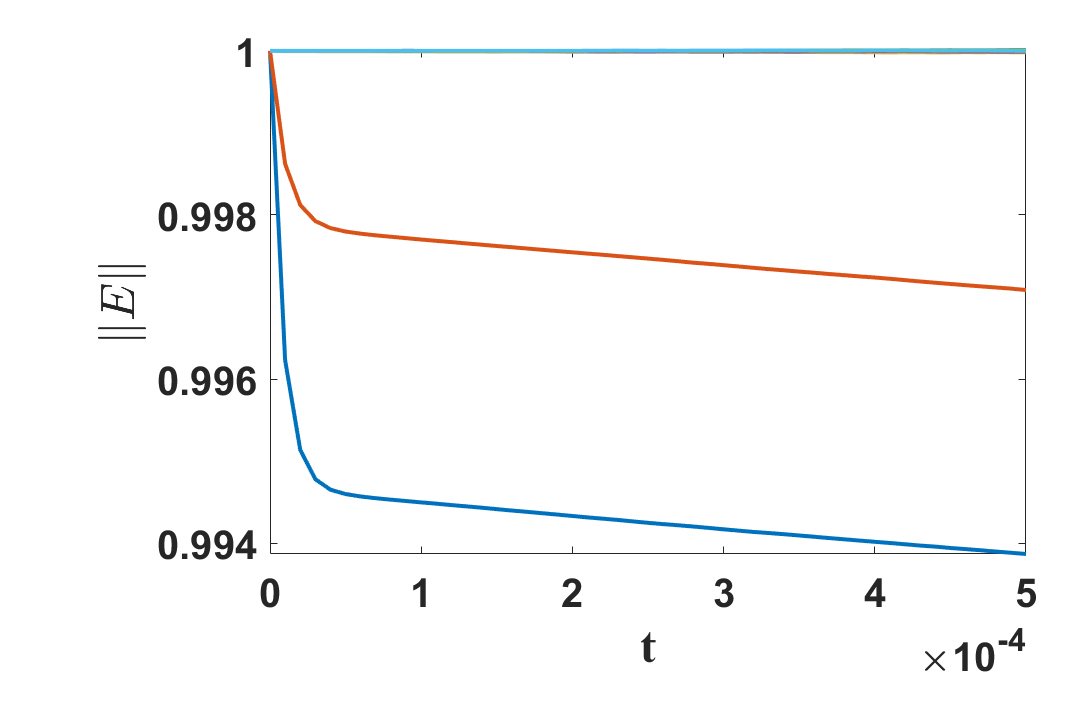}
}
\mbox{
 \hspace{0.45cm}
 \makebox[0.45\textwidth]{(c)}
  \hspace{0.05\textwidth}
 \makebox[0.45\textwidth]{(d)}
 }
 
\caption{1D diffusion of a sine wave using DSEM-SL method using one element, $N_s$=$10^{6}$ samples and $\Delta t$=$10^{-5}$. (a) shows the average solution after 50 time steps along with 10 samples when $P$=8 and $H$=1. The $L^2$ error norm, $\|e\|_{L^2}$, the mass norm, $\|M\|$, and the energy norm, $\|E\|$ are plotted versus time, $t$, in subfigures (b), (c) and (d), respectively.} 
\label{fig:1d_sine_time_evol}
\end{figure}
Figure \ref{fig:1d_sine_time_evol} compares the time evolution of the error and conservation norms of the DSEM-SL scheme for different polynomial orders keeping the number of elements and the number of samples fixed with $H$=1 and $N_s$=$10^6$ respectively. The time evolution of the $L^2$ error shows that the error decreases as the polynomial order is increased consistent with exponential convergence in $P$ for even and odd order polynomials separately. The difference in error between odd and even polynomial approximation is a result of the symmetry of the sine function, which favors the even number of interpolating points for polynomials of an odd degree. At a polynomial order $P$=8 the interpolation error is of the same order as the sampling error. The method accurately conserves mass for upto six decimals for all polynomial orders and the mass conservation improves as the polynomial order increases. The energy norm evolution shows that for low polynomial orders there is a small loss in energy, which reduces for increasing polynomial orders.

\begin{figure} 
\centering 
\mbox{ \includegraphics[width=0.5\textwidth]{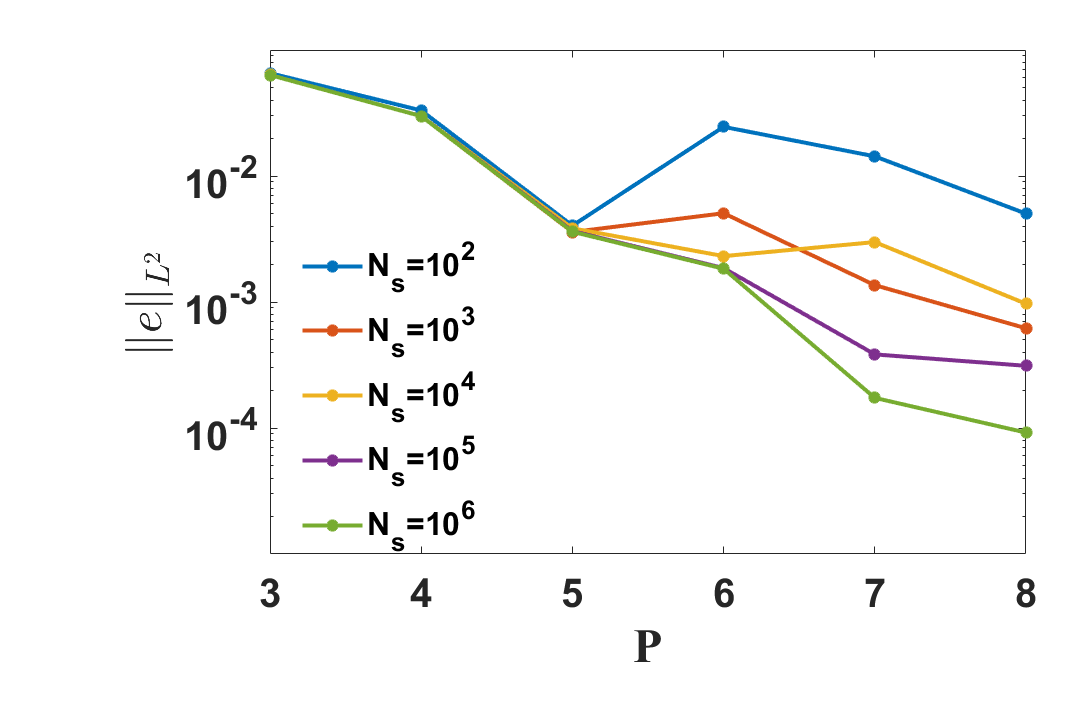}
       \includegraphics[width=0.5\textwidth]{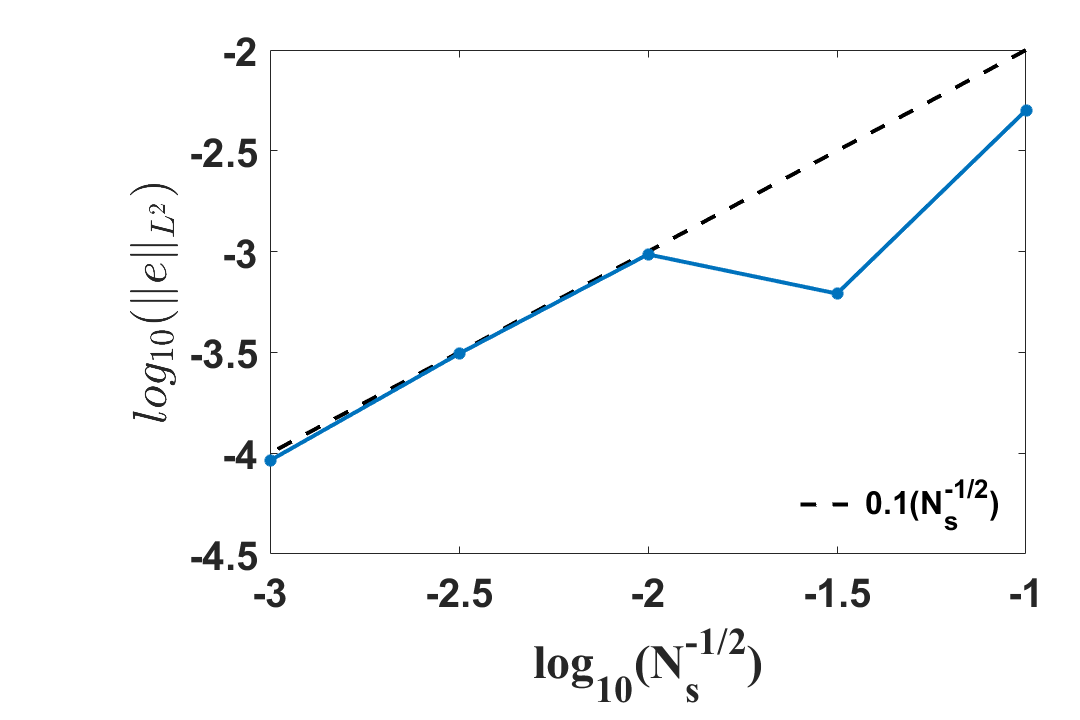}
      }
\mbox{
 \hspace{0.45cm}
 \makebox[0.45\textwidth]{(a)}
 \hspace{0.05\textwidth}
 \makebox[0.45\textwidth]{(b)}
 }
 
\caption{1D diffusion of a sine wave using DSEM-SL method using one element and $\Delta t$=$10^{-5}$. (a) plots the $P$ convergence of the $L^2$ error for different number of samples, $N_s$. (b) plots the $N_s$ convergence of the error when $P$=8.} 
\label{fig:1d_sine_pconv}
\end{figure}
 Figure \ref{fig:1d_sine_pconv}a illustrates the effect of the number of samples on the $P$ convergence.  The sampling error, which can be expected to be on the order of  $N_s^{-1/2}$, is found to be approximately $\mathcal{O}(10^{-4})$ for $N_s$=$10^6$ and is similar to the spatial approximation error for $P$ = 8.
 With a reduced number of samples $N_s$ the error at $P$=8 increases consistently according
 to the sampling convergence. This is confirmed by the linear trend of the error versus $\log_{10}({N_s^{-1/2})}$in Figure \ref{fig:1d_sine_pconv}b.

\begin{figure} 
\centering 
\mbox{ \includegraphics[width=0.5\textwidth]{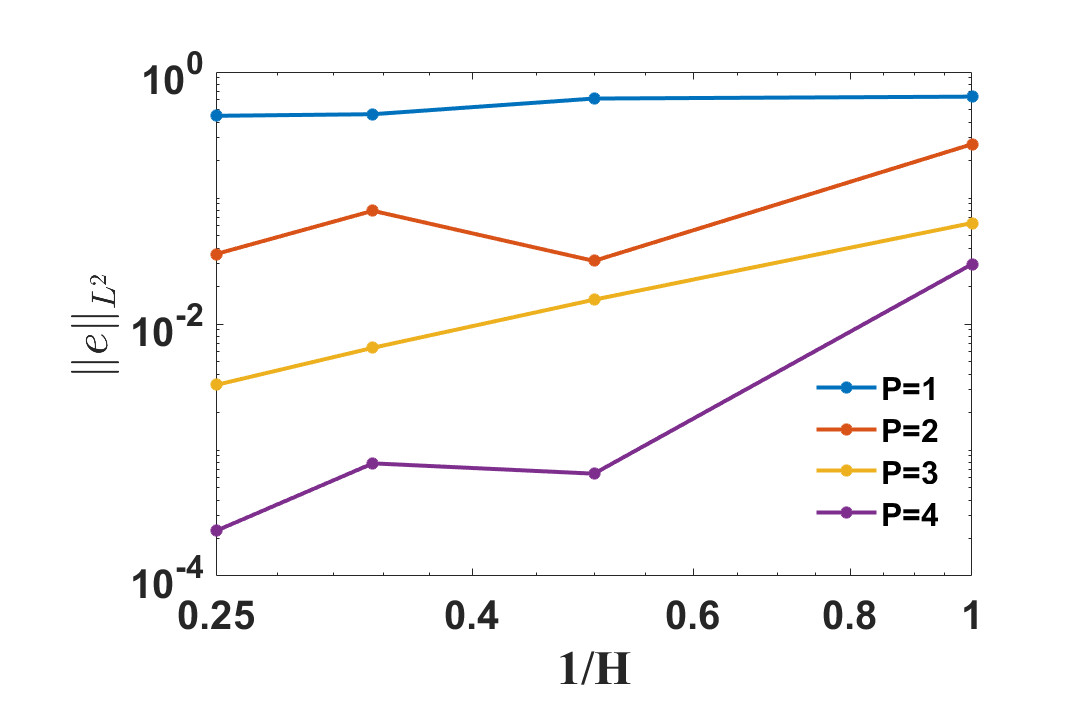}
      }
 
\caption{1D diffusion of a sine wave using DSEM-SL method using $10^6$ samples and $\Delta t$=$10^{-5}$. Plot of the $H$ convergence of the $L^2$ error for different polynomial orders.} 
\label{fig:1d_sine_hconv_old}
\end{figure}
\begin{figure} 
\centering 
\mbox{ 
   \includegraphics[width=0.5\textwidth]{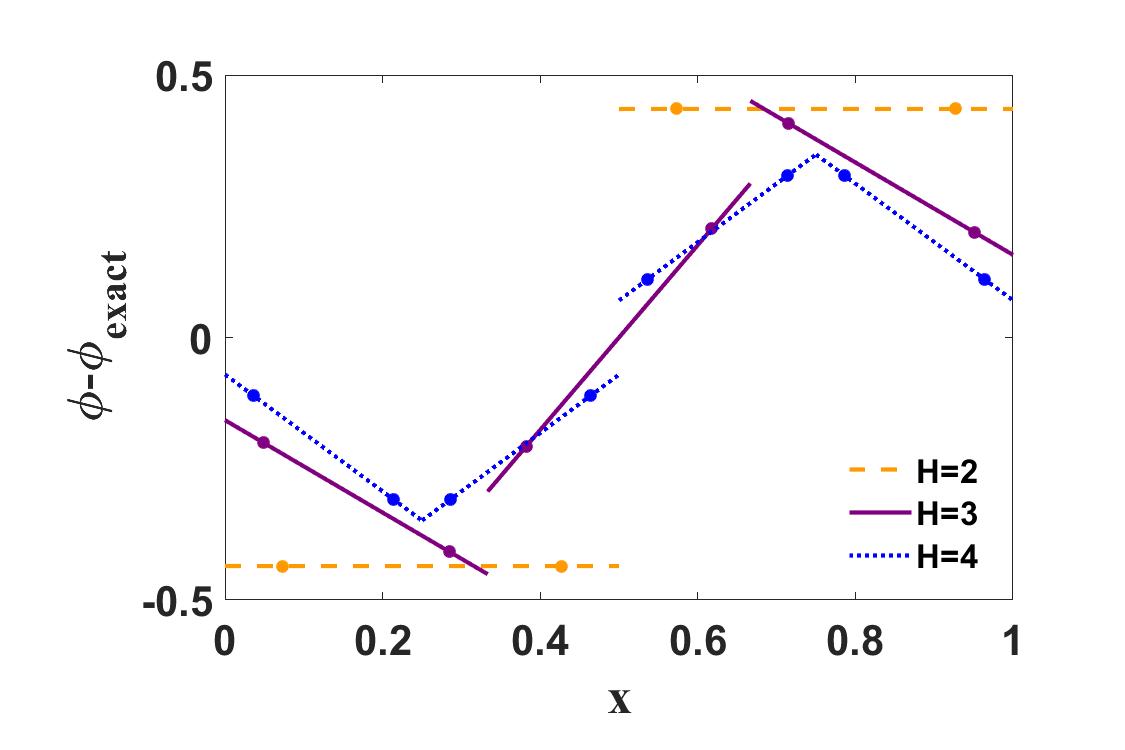}
  \includegraphics[width=0.5\textwidth]{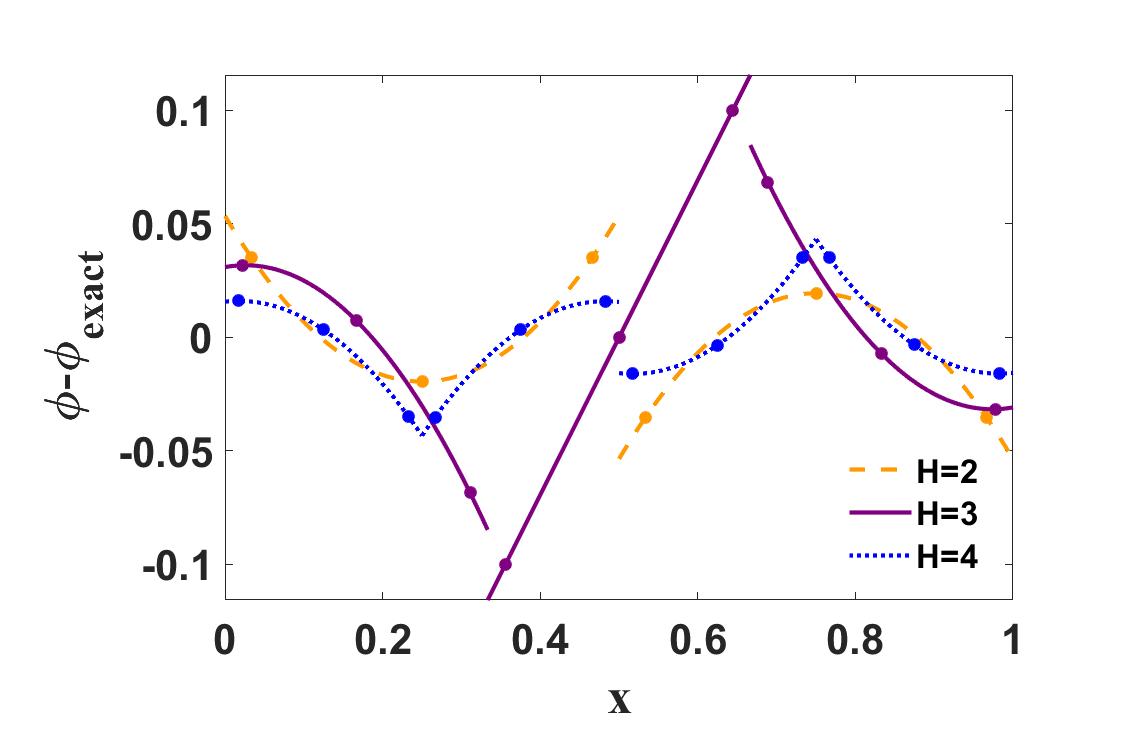} 
}
\mbox{
 \hspace{0.45cm}
 \makebox[0.45\textwidth]{(a)}
  \hspace{0.05\textwidth}
 \makebox[0.45\textwidth]{(b)}
 }
\centering 
\mbox{ 
   \includegraphics[width=0.5\textwidth]{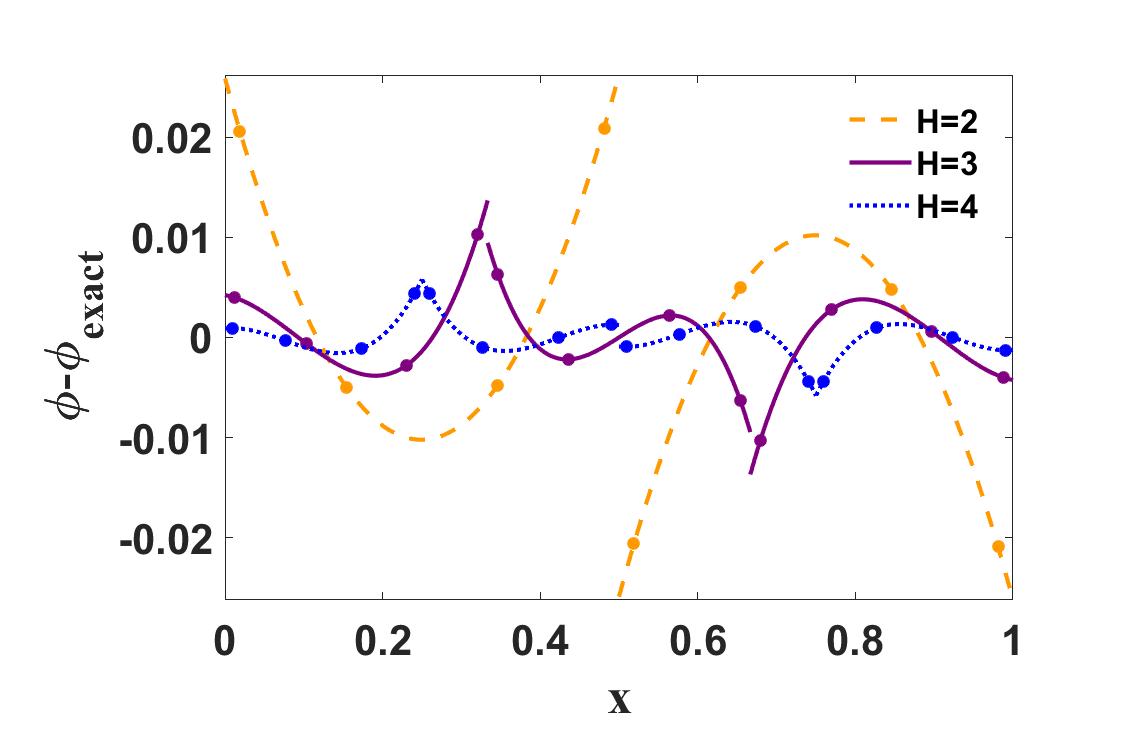}
    \includegraphics[width=0.5\textwidth]{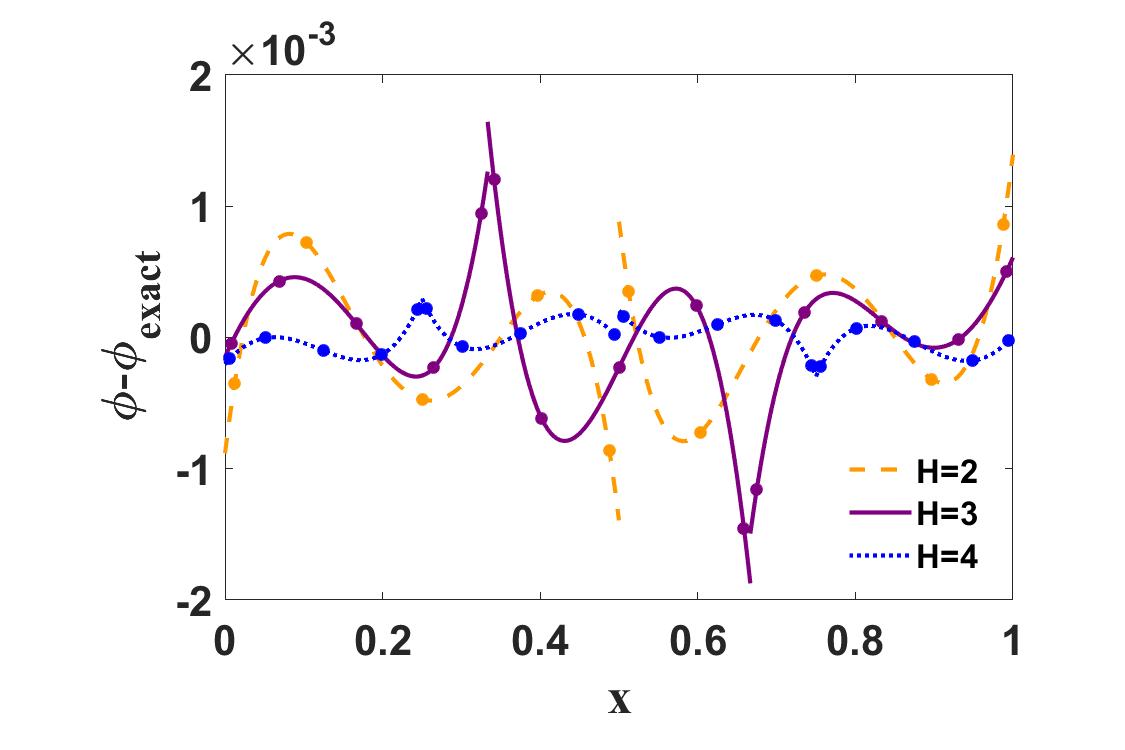}
}
\mbox{
 \hspace{0.45cm}
 \makebox[0.45\textwidth]{(c)}
   \hspace{0.05\textwidth}
 \makebox[0.45\textwidth]{(d)}
 }
 
\caption{1D diffusion of a sine wave using DSEM-SL method using $10^6$ samples and $\Delta t$=$10^{-5}$. Plot of the local error for different number of elements. (a) $P$=1, (b) $P$=2, (c) $P$=3 and (d) $P$=4. } 
\label{fig:1d_sine_local_error}
\end{figure}

\begin{figure} 
\centering 
\mbox{ \includegraphics[width=0.5\textwidth]{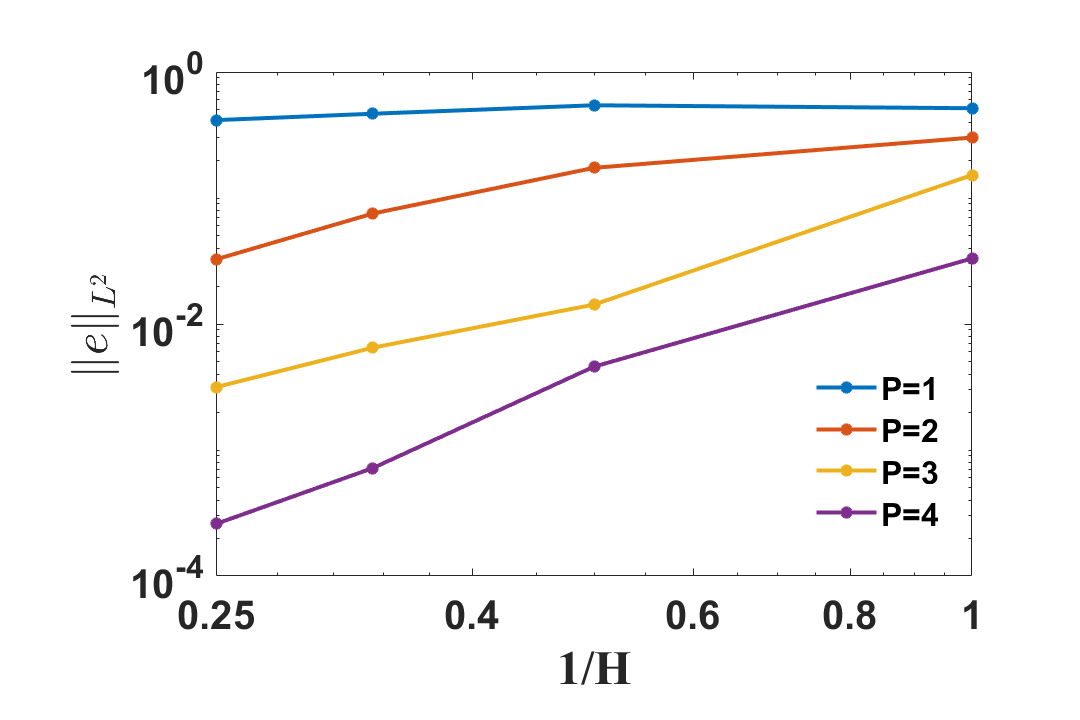}
      }
 
\caption{1D diffusion of a sine wave using DSEM-SL method using $10^6$ samples and $\Delta t$=$10^{-5}$. Initial condition used $\phi(x,0)$ =$\sin(2\pi (x-0.1))+2$. Plot of the $H$ convergence of the $L^2$ error for different polynomial orders.} 
\label{fig:1d_sine_hconv}
\end{figure}

A log-log plot of the error versus the grid spacing $h=1/H$ in Figure \ref{fig:1d_sine_hconv_old} using $10^6$ samples is linear and show that the methods converges in an algebraic manner according to $\mathcal{O}(h^p)$. It can be observed that the error convergence plot when $P$=2 and $P$=4 shows anomaly for three elements $H$=3. The local error is plotted in Figure \ref{fig:1d_sine_local_error} shows that for three elements and even polynomial order, a quadrature point is located exactly at the center location of the sine wave. The approximation in the center element then yields overshoots at the edges of the center elements and the erratic convergence behavior.
To avoid this behaviour, the $h$-convergence was computed after shifting the $\sin$ function. A log-log plot of the error versus the grid spacing is shown in Figure \ref{fig:1d_sine_hconv} using an initial condition of $\phi(x,0)$ =$\sin(2\pi (x-0.1))+2$. The plot shows the expected algebraic convergence trend without anomalies.   

\begin{figure} 
\centering 
\mbox{ \includegraphics[width=0.5\textwidth]{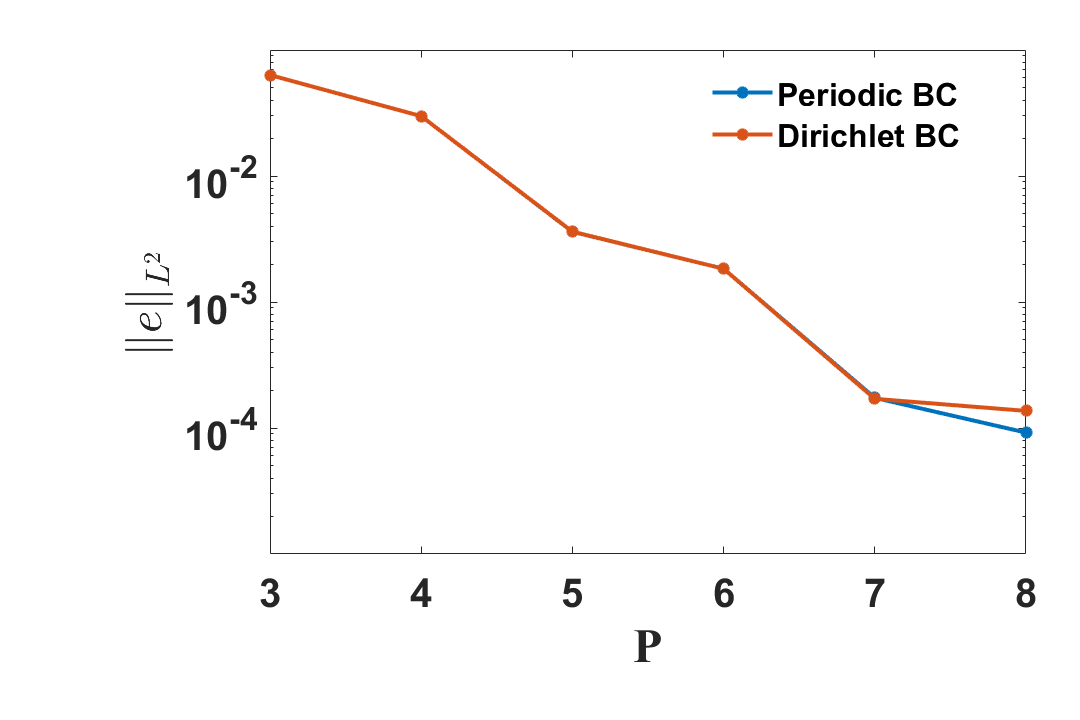}
      }
\caption{Comparing of the log-linear plot of the $L^2$ error
versus $P$ for a one dimensional diffusion of a sine wave using DSEM-SL method 
with periodic and Dirichlet boundary conditions. One element is considered and $\Delta t$=$10^{-5}$} 
\label{fig:1d_sine_pconv_BC}
\end{figure}

A computation that employs a Dirichlet boundary condition shows no discernible differences with one that uses periodic boundary conditions (Figure \ref{fig:1d_sine_pconv_BC}).
  
\label{sec:1d_sine}
\subsection{Discussion on sampling: Realistic simulation of the sine wave}
\begin{figure} 
\centering 
\mbox{ \includegraphics[width=0.5\textwidth]{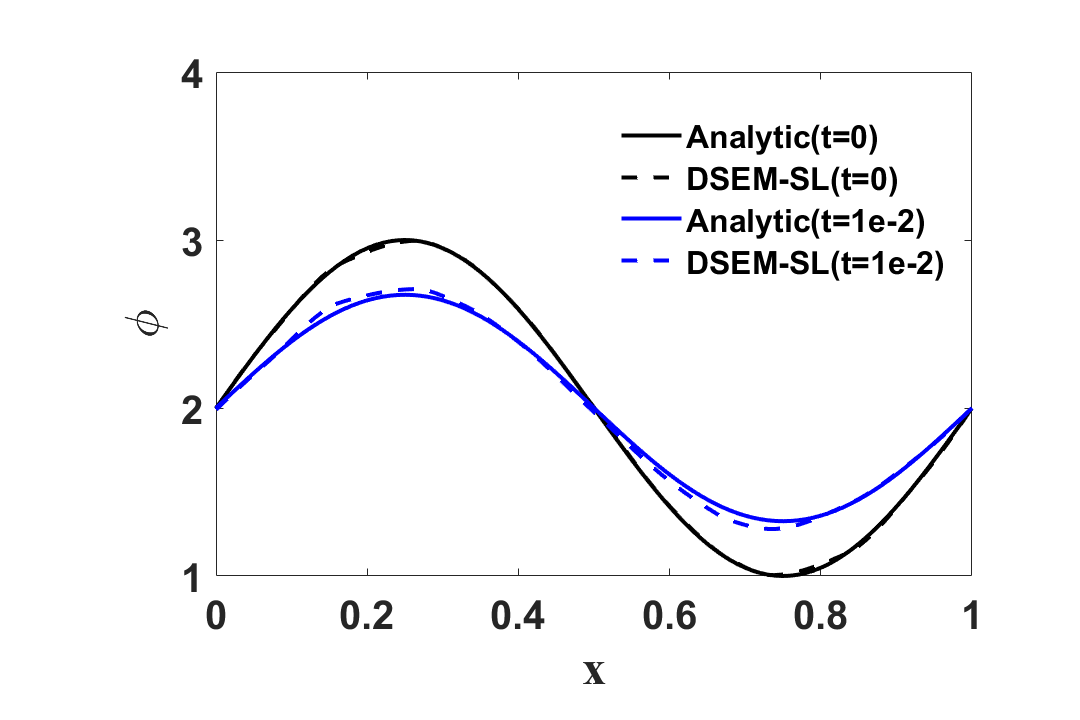}
       \includegraphics[width=0.5\textwidth]{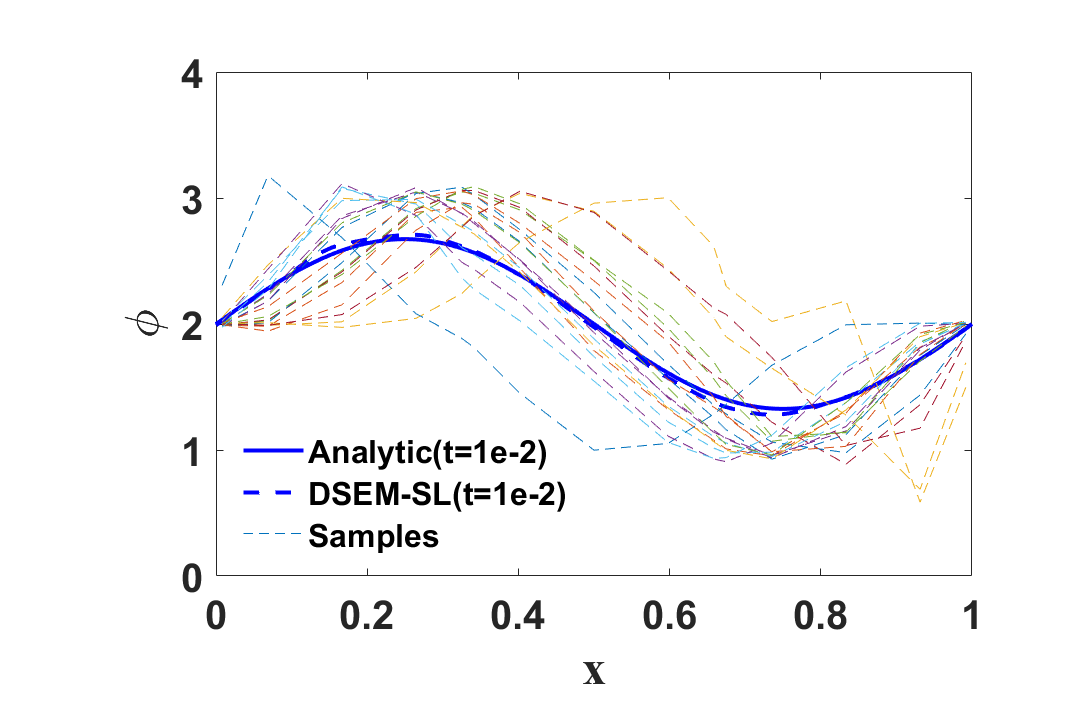}
      }
\mbox{
 \hspace{0.45cm}
 \makebox[0.45\textwidth]{(a)}
 \hspace{0.05\textwidth}
 \makebox[0.45\textwidth]{(b)}
 }
 
\caption{a) shows the plot comparing the DSEM-SL method with the analytic solution for a one dimensional Sine function. DSEM-SL method is run using $H$=3, $P$=4 using $N_s$=100 samples and the the averaged solution is plotted against the analytic solution for $t$=0 and $t$=1e-2. b) shows that when using Dirichlet boundary condition, some samples develop high gradient at the boundaries after a long period of time. } 
\label{fig:1d_sine_realistic}
\end{figure}

In practical simulation of more complex problems over longer times, the computational burden to generate $N_s$=$10^6$ samples is too high for current day computational resources. Typically in engineering computations fewer samples are used per point on the order of tens to hundreds,
yielding sampling errors of a few percent.
In Figure \ref{fig:1d_sine_realistic} (a), we illustrate the performance of the DSEM-SL method of the diffused sine wave generated with a hundred samples,  $N_s$=100.  Over a time span of $t$=1e-2, the amplitude of the sine wave has reduced significantly, a measure for the diffusion. Per the expectation and comparable to Lagrangian methods, the semi-Lagrangian solution is in good comparison with the analytical solution within a few percentages accuracy. 
 Figure \ref{fig:1d_sine_realistic} (b) shows that when Dirichlet boundary conditions are used in longer time simulations, some samples develop high gradient at the boundaries. To prevent the samples from developing high gradients at the boundaries, we need to re-seed the samples from the average solution after every few time steps. In this example we re-seed the samples after every 100 time steps.    
\subsection{Comparison with random walk methods: One dimensional Sine function}

To assess the performance of DSEM-SL in relation to exisiting methods,
we compare it to the strong random walk (RW) method, weak random walk method and the generalized  random
walk (GRW) method as described in Section \ref{sec:prev_methods}. For constant diffusion of a sine function,
we focus on spatial accuracy and its convergence.
To do so, we ensure that the sampling error and the time integration error is kept low by using $10^6$ samples and a time step of  $10^{-5}$. Because of the large sample rate, the simulations are computationally intensive and
we compute 50 time steps only.  The short simulation times lead to a lower limit on the number of grid points for the weak RW and GRW methods to $N$=31 since $\Delta x \leq \sqrt{2 \Delta t}$. 

\begin{figure} 
\centering 
\mbox{ \includegraphics[width=0.5\textwidth]{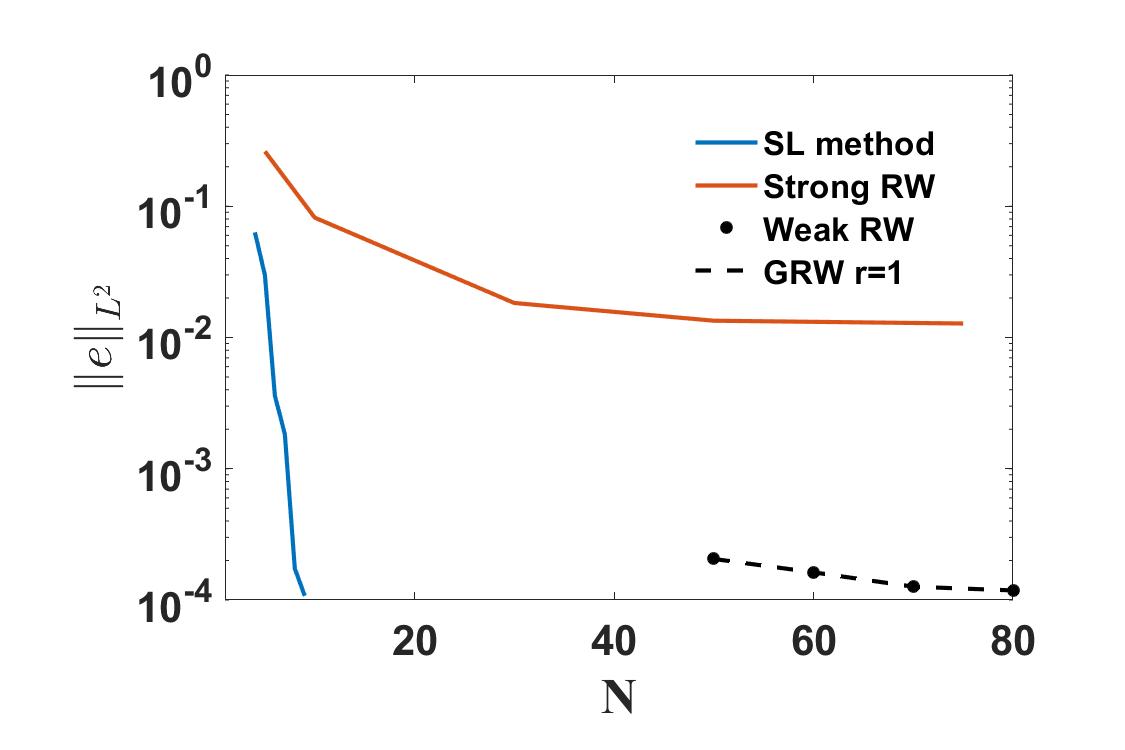}
      }
 
\caption{1D diffusion of a sine wave comparing DSEM-SL method, strong RW method, Weak RW method and GRW method using $10^6$ samples. (a) plots the $L^2$ error vs the number of grid points, $N$.} 
\label{fig:1d_sine_compare}
\end{figure}

After 50 time steps, the number of grid points $N$ is plotted versus the $\|e\|_{L^2}$ error norm  in Figure \ref{fig:1d_sine_compare}. The DSEM-SL method shows exponential convergence whereas  the strong RW, weak RW and the GRW methods have algebraic convergence only. The DSEM-SL method method can achieve an error of around $10^{-4}$ using up to five times fewer number of points compared to the GRW method. 

\subsection{One dimensional constant diffusion: Gaussian function}

Because the diffusion of the sine wave displayed some odd convergence behaviors that are directly related
to symmetries in the polynomial point distribution and sine function behavior,
we test another pure diffusion case  with the diffusion coefficient, $D$=1 for a different initial condition. 
In a domain $x$=$[-1,1]$, we set the initial condition as
a Gaussian function according to an analytical 
solution of (\ref{eq:fokkerplanck})  as
\begin{equation}
 \phi(x,t) = {1 \over \sqrt{4\pi t}} \exp{\left({-x^2 \over 4t}\right)},
\end{equation}
 at $t=t_0$=$0.05$.
 Dirichlet boundary conditions are specified according to the analytical solution.
The time step is set to $\Delta t$= $10^{-5}$ and simulations are carried out for 100 time steps.

\begin{figure} 
\centering 
\mbox{ 
   \includegraphics[width=0.5\textwidth]{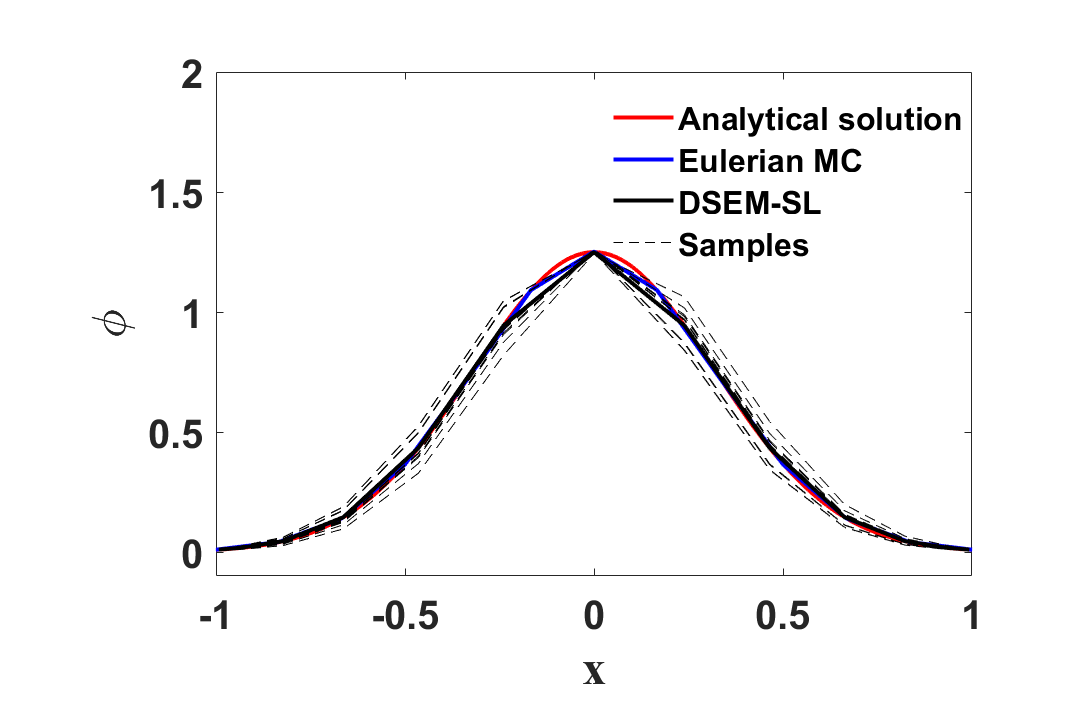}
  \includegraphics[width=0.5\textwidth]{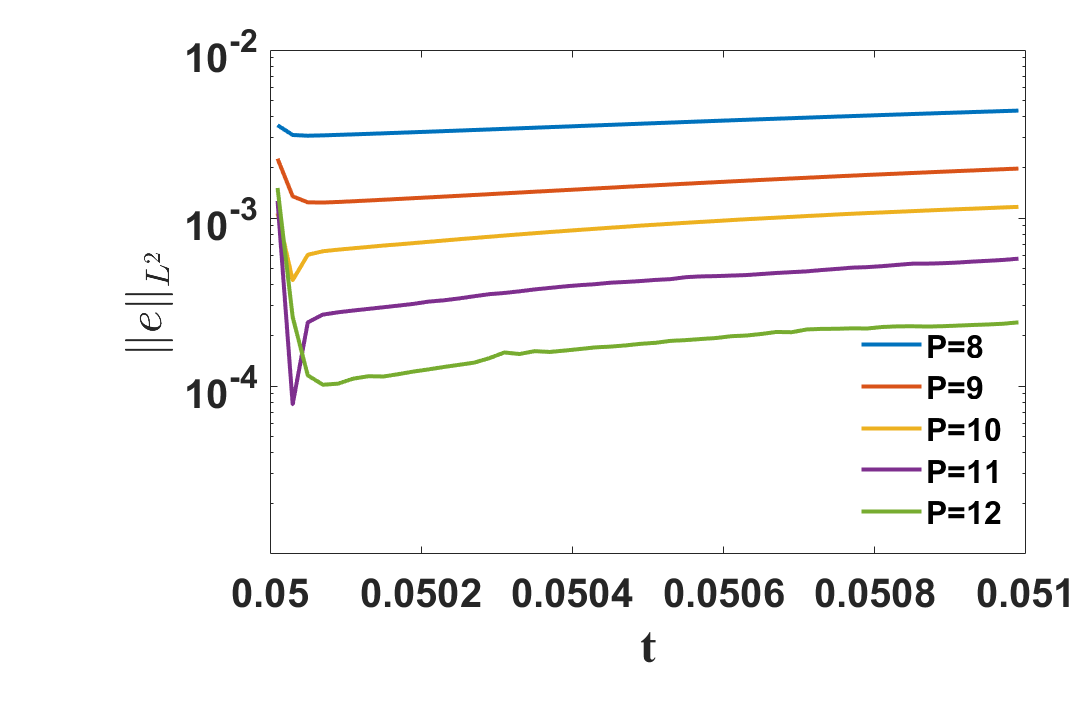} 
}
\mbox{
 \hspace{0.45cm}
 \makebox[0.45\textwidth]{(a)}
  \hspace{0.05\textwidth}
 \makebox[0.45\textwidth]{(b)}
 }
\centering 
\mbox{ 
   \includegraphics[width=0.5\textwidth]{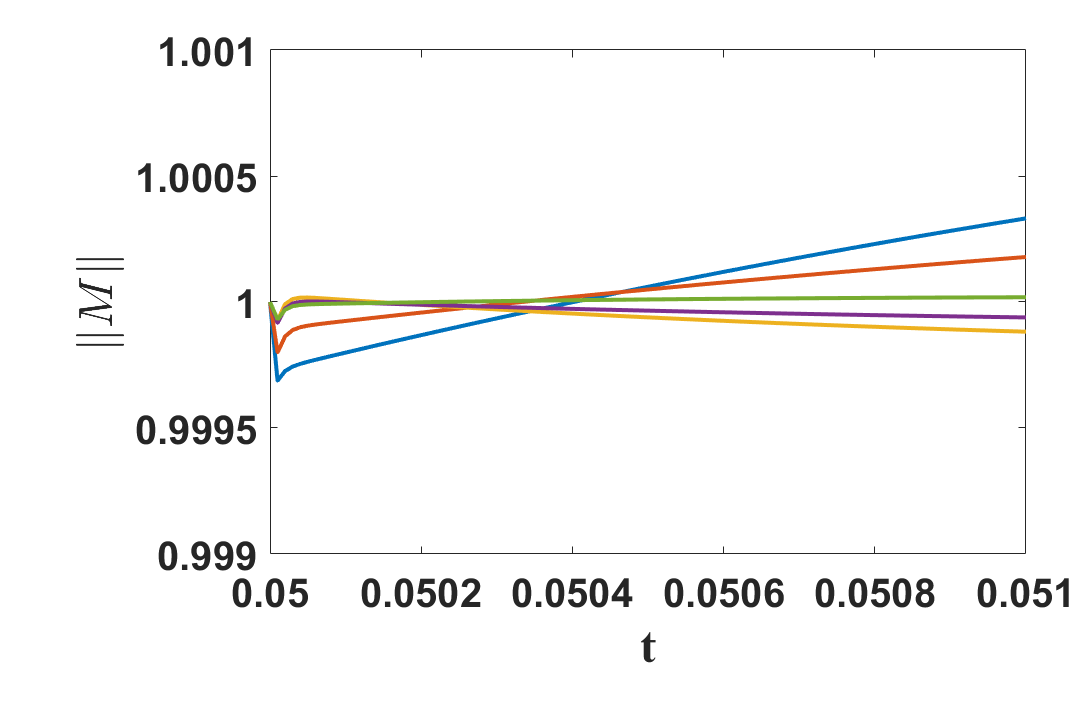}
    \includegraphics[width=0.5\textwidth]{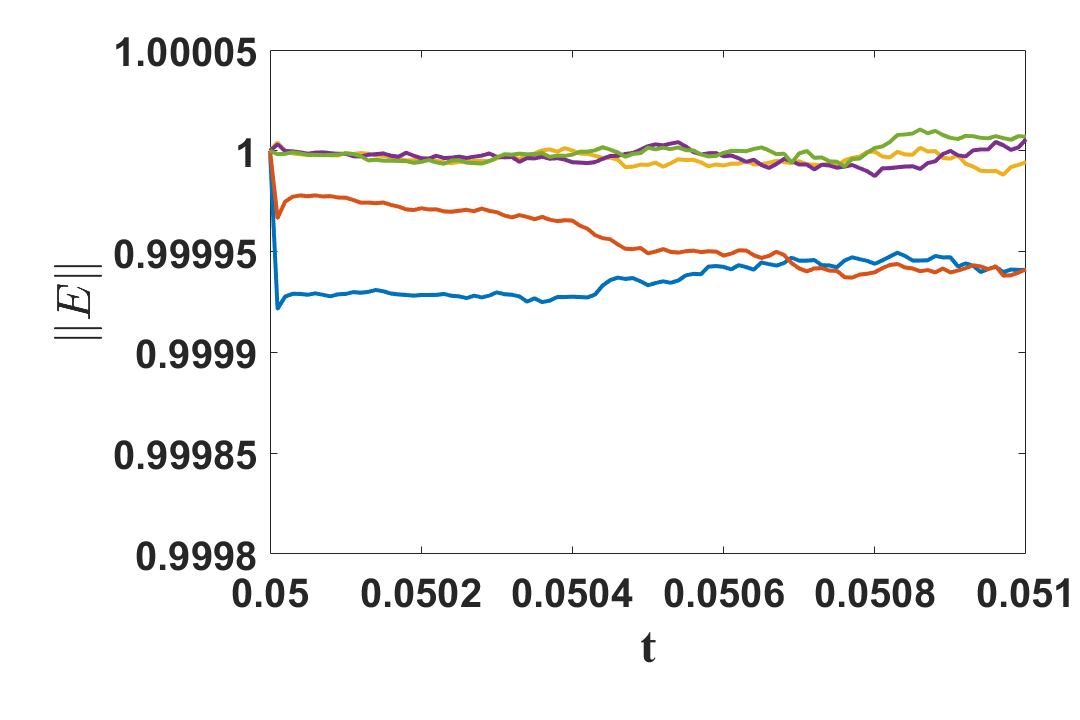}
}
\mbox{
 \hspace{0.45cm}
 \makebox[0.45\textwidth]{(c)}
   \hspace{0.05\textwidth}
 \makebox[0.45\textwidth]{(d)}
 }
 
\caption{1D diffusion of a Gaussian function using DSEM-SL method using one element, $N_s$=$10^{6}$ samples and $\Delta t$=$10^{-5}$. (a) shows the averaged solution and 10 samples after 100 time steps for $P$=12 and $H$=1. A second-order Eulerian Monte Carlo solution with similar spatial resolution is shown for comparison. The $L^2$ error norm, $\|e\|_{L^2}$, the mass norm, $\|M\|$, and the energy norm, $\|E\|$ are plotted versus time, $t$, in subfigures (b), (c) and (d), respectively.} 
\label{fig:1d_Gaussian_time_evol}
\end{figure}
\begin{figure} 
\centering 
\mbox{ \includegraphics[width=0.5\textwidth]{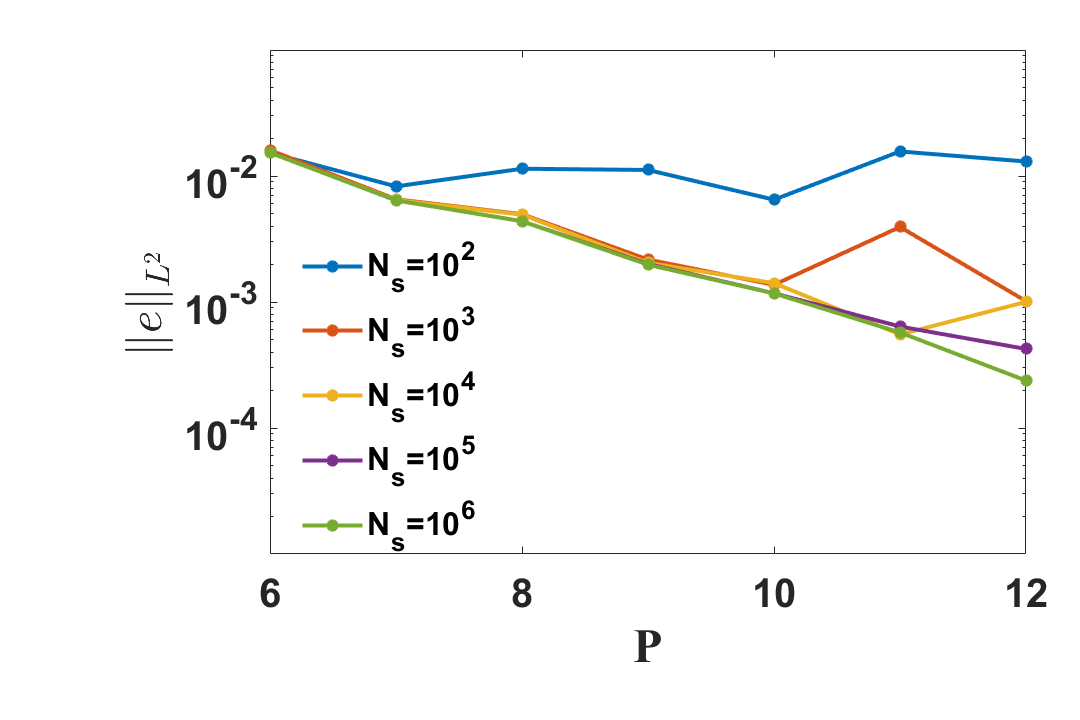}
       \includegraphics[width=0.5\textwidth]{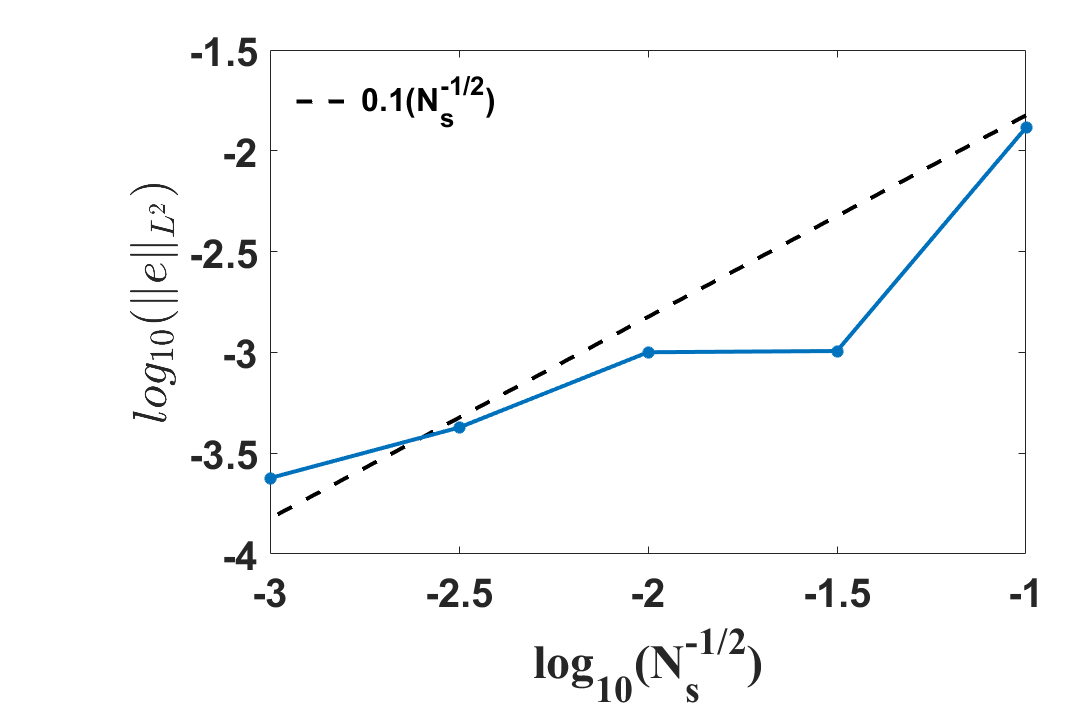}
      }
\mbox{
 \hspace{0.45cm}
 \makebox[0.45\textwidth]{(a)}
 \hspace{0.05\textwidth}
 \makebox[0.45\textwidth]{(b)}
 }
 
\caption{1D diffusion of a Gaussian function using DSEM-SL method using one element and $\Delta t$=$10^{-5}$. (a) plots the $P$ convergence of the $L^2$ error for different number of samples, $N_s$. (b) plots the $N_s$ convergence of the error when $P$=12.} 
\label{fig:1d_Gaussian_pconv}
\end{figure}
\begin{figure} 
\centering 
\mbox{ \includegraphics[width=0.5\textwidth]{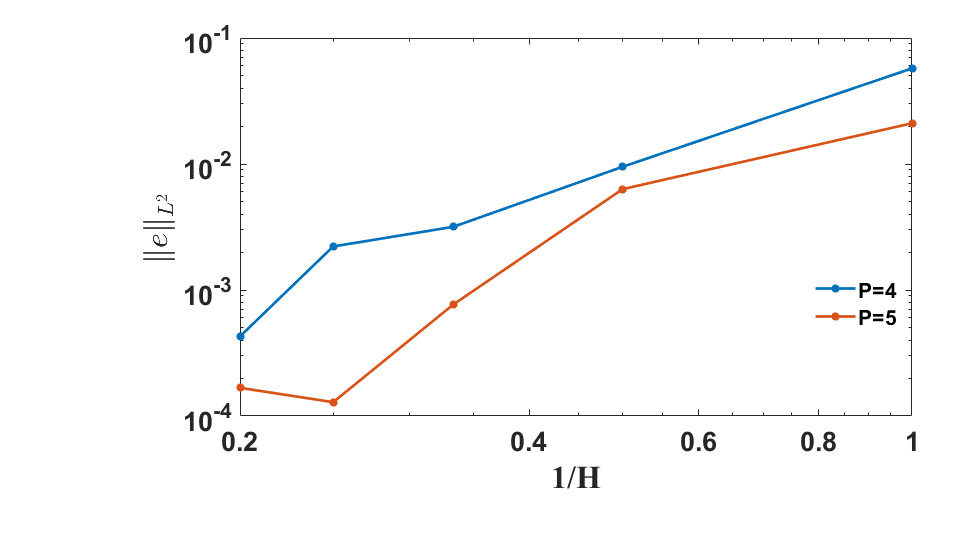}
      }
 
\caption{1D diffusion of a Gaussian function using DSEM-SL method using $10^6$ samples and $\Delta t$=$10^{-5}$. Plot of the $H$ convergence of the $L^2$ error for different polynomial orders.} 
\label{fig:1d_Gaussian_hconv}
\end{figure}

Figure \ref{fig:1d_Gaussian_time_evol} compares the time evolution of the conservation properties and the error of the DSEM-SL scheme for different polynomial orders keeping the number of elements and the number of samples fixed with $H$=1 and $N_s$=$10^6$ respectively. Figure \ref{fig:1d_Gaussian_time_evol}b shows that the $\|e\|_{L^2}$ error decreases as the polynomial order increases. Figure \ref{fig:1d_Gaussian_time_evol}c and d indicate mass and energy conservation upto four decimal places. The plot of the $P$ convergence of the $L^2$ error for different number of samples, $N_s$ is shown in Figure \ref{fig:1d_Gaussian_pconv}a. The $L_2$ error at $P$=12 for different $N_s$ values is plotted in Figure \ref{fig:1d_Gaussian_pconv}b. Similar to the 1D sine wave test case, the $P$ convergence curves follow the expected spectral convergence until the polynomial interpolation error is of the same order of the sampling error and the convergence of the sampling error follows the expected trend with a slope of $1/\sqrt{N_s}$. The difference however with the sine wave test case is that the error behavior vs even/odd polynomial orders for $H=1$ is not noticeable anymore, which confirms that this behavior
is specific to the sine case. If an odd number of points is used to approximate the sine function, then the middle point always has the exact value. This is not the case for the Gaussian initial condition.
The $h$-convergence in the $\|e\|_{L^2}$ error is shown in Figure \ref{fig:1d_Gaussian_hconv}. The plot shows an expected algebraic convergence in the error as the number of elements is increased from $H$=1 to $H$=5 using $N_s$=$10^6$ samples for polynomial orders $P$=4 and $P$=5.

\subsection{One dimensional Ornstein–Uhlenbeck test case}
To test the scheme for formulations that involve  both advection and diffusion physics, we consider the analytical Ornstein–Uhlenbeck solution \cite{OrnsteinUhlenbeck30} for the probability density function, $\Pdf(x,t)$, in the Fokker-Planck equation (\ref{eq:fokkerplanck}).
The analytical solution for the Ornstein-Uhlenbeck process with $u=-\alpha x$ is given by, 
\begin{equation} \label{eq:advectiondiffusion}
\Pdf(x,t)=  \sqrt{{\alpha \over 2 \pi D (1-\exp{(-2 \alpha \tau)})}} \exp{\left[-{\alpha \over 2D} {{(x-x_0\exp{(-\alpha \tau)})^2} \over { (1-\exp{(-2\alpha \tau)})}}\right]}, 
\end{equation}
where, $\tau$=$t-t_0$, $t_0$ is the initial time. In this test we take $\alpha$=1, $D$=1,  $t_0$=0.25 and  $x_0=2$.


In order to use the DSEM-SL method, we re-write the 1D version of (\ref{eq:fokkerplanckPhi}) with the specific source term $S(\psi;x,t)=-\psi \frac{\partial u}{\partial x} $,

\begin{equation}\label{eq:6.5_1}
\frac{\partial \Pdf_\phi}{\partial t} + u \frac{\partial \Pdf_\phi}{\partial x} = \frac{\partial}{\partial x}\left( D \frac{\partial \Pdf_\phi}{\partial x} \right) + \frac{\partial}{\partial \psi} \left[ \psi \frac{\partial u}{\partial x} \Pdf_\phi \right]
\end{equation}

\noindent Taking the first moment $\int \psi \cdot d\psi$, of all terms and applying integration by parts to the last term on the RHS, we get

\begin{equation}\label{eq:6.5_2}
\frac{\partial \phi}{\partial t} + u \frac{\partial \phi}{\partial x} = \frac{\partial}{\partial x}\left( D \frac{\partial \phi}{\partial x} \right) - \frac{\partial u}{\partial x} \phi,
\end{equation}

\noindent which is equivalent to (\ref{eq:conv-diffusionPrecursor}) for constant $D$. We solve the equivalent SDE for the particles position and the transport equation for $\phi^*$ along the particles' trajectories,
\begin{eqnarray}
d {X_t} &=& {u}({X}_t,t) dt + \sqrt{2D}({X}_t,t) dW_t, \\
\frac{D \phi^*}{D t} &=& -\phi^* \frac{\partial u }{\partial x}.
\end{eqnarray}
 The simulations are initialized according to the analytical solution at $t$=0.25 and are carried out in a domain $x \in [-4, 6]$ using $N_s$=100 samples and a polynomial order $P$=17. The samples are re-seeded every 100 time steps to prevent high gradients appearing near the boundaries.

\begin{figure} 
\centering 
\mbox{ \includegraphics[width=0.5\textwidth]{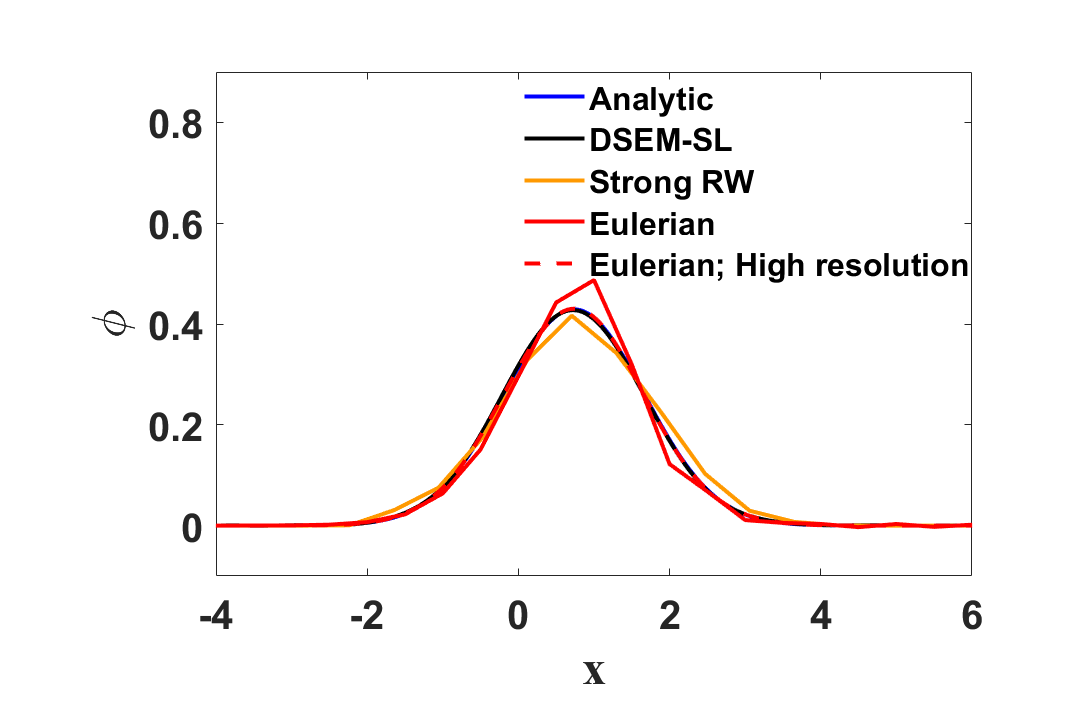}
  \includegraphics[width=0.5\textwidth]{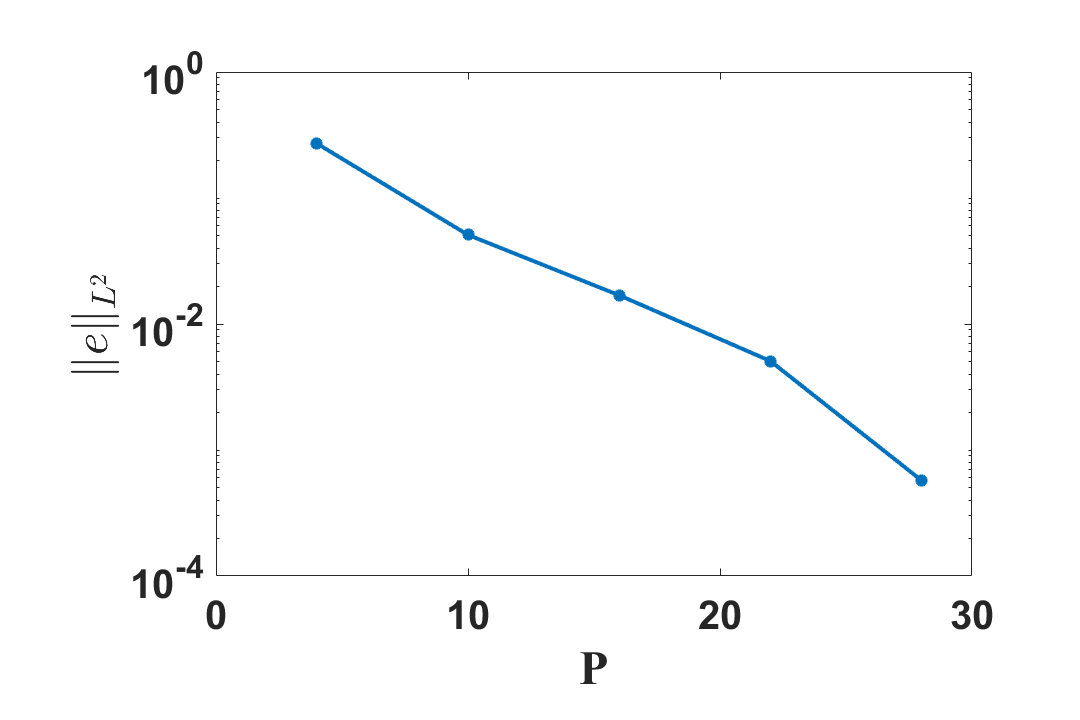}
      }
\mbox{
 \hspace{0.45cm}
 \makebox[0.45\textwidth]{(a)}
  \hspace{0.05\textwidth}
 \makebox[0.45\textwidth]{(b)}
 }
 
\caption{DSEM-SL method for a one dimensional Ornstein–Uhlenbeck process. (a) plots the solution at initial time, $t$=0.25 and at $t$=1, the solution is plotted against the analytic solution and a strong RW method. DSEM-SL is run using $H$=1 with $N_s$=100 samples and $P$=17. (b) plots the $P$-convergence in $\|e\|_{L^2}$ error when $H$=1 and $N_s$=$10^4$.} 
\label{fig:advectiondiffusion}
\end{figure}

Figure \ref{fig:advectiondiffusion}a plots the solution of the DSEM-SL method vs the analytical solution, the strong RW method and the EMC method (implemented with a second-order central differencing scheme) at $t$=1. For the latter, we provide results at two levels of resolution: a grid with the same spatial resolution as the DSEM-SL scheme, and a finer grid with $16$ times more points. The strong RW method uses the same number grid points and samples as the DSEM-SL method; $N$=$18$ and $N_s$=$100$ samples. The solution using DSEM-SL method matches the analytical solution better and is smoother compared to both the strong RW method and the EMC method with the same particle resolution. The higher resolution EMC solution has accuracy similar to that of DSEM-SL, but requires $16$ times more points. 

Figure \ref{fig:advectiondiffusion}b shows the $\|e\|_{L^2}$ error convergence on polynomial order using $H$=1 and $N_s$=$10^{4}$. The $P$ convergence is observed to show spectral convergence for polynomial orders from $P$=$4$ to $P$=$28$.    


\subsection{One dimensional non-constant diffusion: Sine function}
Next we test the DSEM-SL algorithm for a non-constant diffusion coefficient of $D$=$x^2$ and $u=0$, for which Equation (\ref{eq:fokkerplanckPhi}) can then be written as, 
\begin{eqnarray}\label{eq:6.6_1}
    {\partial{\Pdf_\phi} \over \partial t} &=& {\partial \over  \partial x} \left( x^2 \frac{\partial \Pdf_\phi}{\partial x} \right).
\end{eqnarray}

To show equivalence with (\ref{eq:fokkerplanck}), we compare to a strong RW solution with
\begin{eqnarray}
d {X_t} &=& 2 X_t dt + \sqrt{2 {{X_t}}^2} dW_t \\
d \phi^* &=& 0,
\end{eqnarray}

\noindent for which the Fokker-Planck equation is

\begin{eqnarray}\label{eq:6.6_2}
    {\partial{\Pdf} \over \partial t} + \frac{\partial}{\partial x} \left( 2x \Pdf \right) &=& {\partial^2 \over  \partial x^2} \left( x^2 \Pdf \right) \nonumber \\
    &=& \frac{\partial}{\partial x}\left( x^2 \frac{\partial \Pdf}{\partial x} +2x \Pdf  \right),
\end{eqnarray}

\noindent which, after cancellation of the $\frac{\partial}{\partial x}\left( 2x \Pdf \right)$ term on both sides, has the same functional form (\ref{eq:6.6_1}). Note that making (\ref{eq:6.6_1}) and (\ref{eq:6.6_2}) equivalent requires using different drift terms in the DSEM-SL and strong RW procedures, due to the different form of the diffusive terms in (\ref{eq:fokkerplanckPhi}) and (\ref{eq:fokkerplanck}), respectively. Taking the first moment $\int \psi \cdot d\psi$, of all terms in (\ref{eq:6.6_1}) and (\ref{eq:6.6_2}), both equations yield the same PDE for $\phi(x,t)$,

\begin{equation}
{\partial{\phi} \over \partial t} = {\partial \over  \partial x} \left( x^2 \frac{\partial \phi}{\partial x} \right).
\end{equation}

Note that we added a conserved composition variable to the strong RW solution - this does not change the functional form of (\ref{eq:6.6_2}), but allows us to solve for negative values of $\phi$.

\begin{figure} 
\centering 
\mbox{ 
   \includegraphics[width=0.5\textwidth]{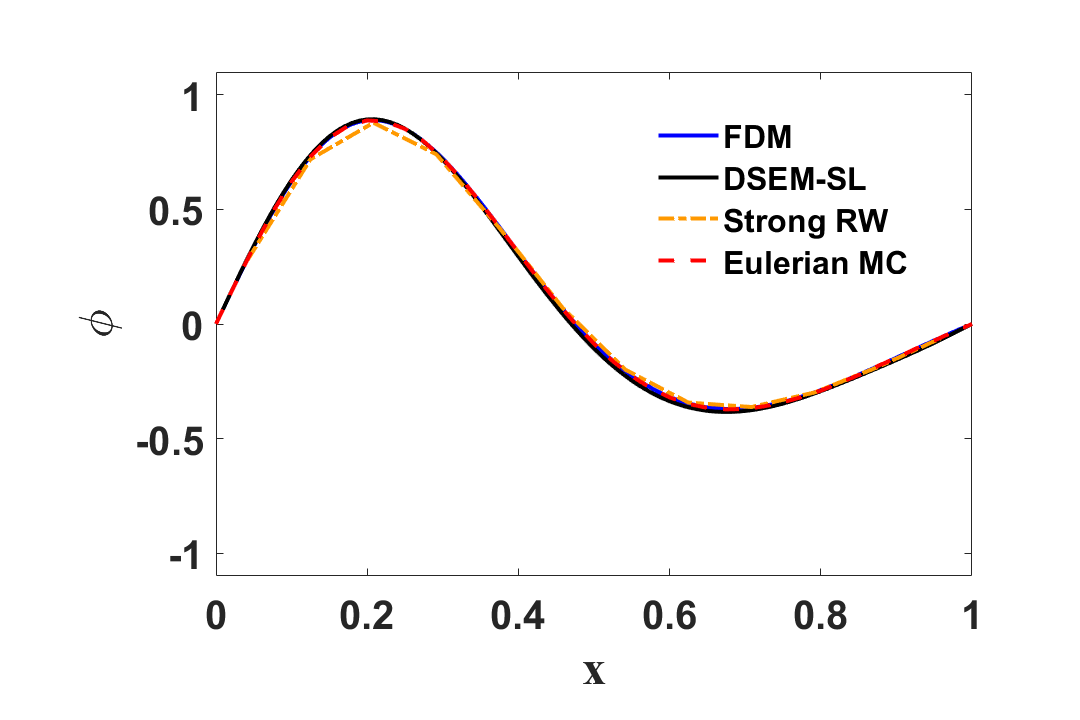}
}
 
\caption{1D non constant diffusion of a sine function: DSEM-SL method $N_s$=$10^{6}$ samples with $P$=11. Plot compares the DSEM-SL solution with the high order finite difference, strong RW and Eulerian Monte Carlo solutions at $t$=$0.05$.} 
\label{fig:1d_sine_nonconstD_soln}
\end{figure}

The DSEM-SL method is solved using $H$=$1$, $P$=$10$ and $N_s$=$100$ with an initial condition, $p(x,0) = \sin(2\pi x)$ in a domain $x \in [0,1]$. A high order finite difference method (FDM) is used as the reference solution. Figure (\ref{fig:1d_sine_nonconstD_soln}) plots the DSEM-SL, FDM, EMC and strong RW solutions at $t$=$0.05$. The DSEM-SL solution matches the high-order FDM solution.  

\subsection{Two dimensional diffusion: Sine function}
The DSEM-SL method extends naturally to multiple dimensions on tensorial grids.
To test, we consider a pure diffusion (drift velocity set to zero) of a tensor product of sine waves in two dimensions.
The initial condition is, $\phi(x,y)$ = $\sin(2\pi x) \sin(2\pi y) +2$ in a domain $x$=$[0,1]$;$y$=$[0,1]$. We set the diffusion coefficient, $D$=$1$. The time step is set to $\Delta t$= $10^{-5}$  and the simulations are carried out for 50 time steps using Dirichlet boundary conditions for different polynomial orders, $P$ and different number of samples, $N_s$. This is the two-dimensional extension of the test case described in \ref{sec:1d_sine}. 

\begin{figure} 
\centering 
\mbox{ 
  \includegraphics[width=0.5\textwidth]{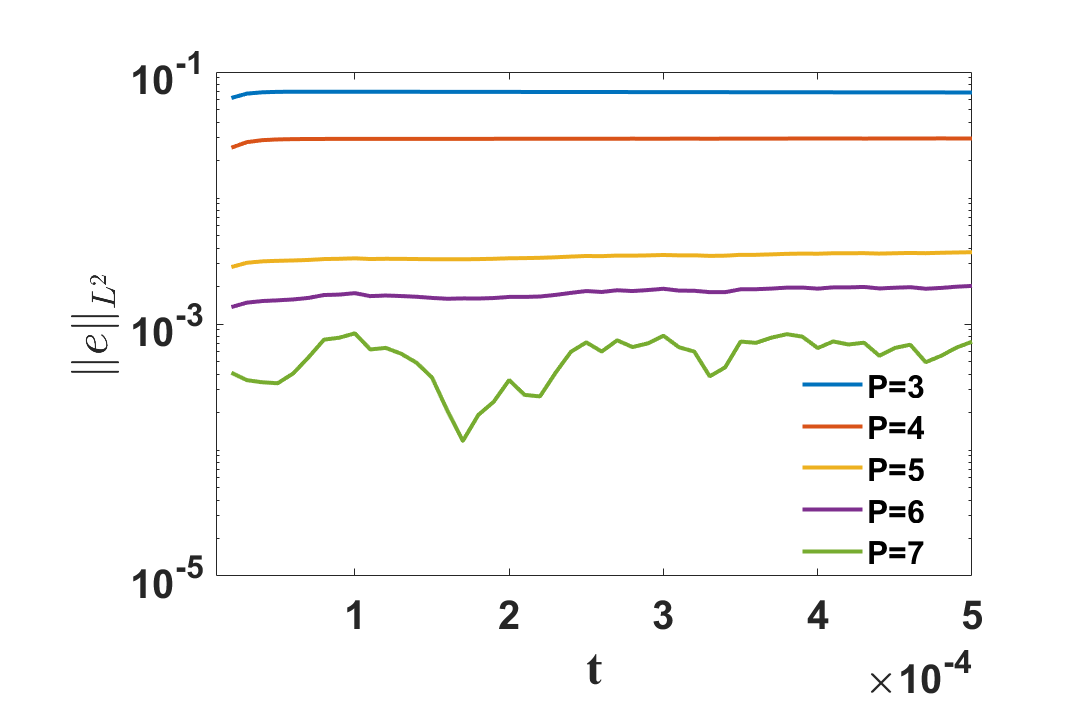}
  \includegraphics[width=0.5\textwidth]{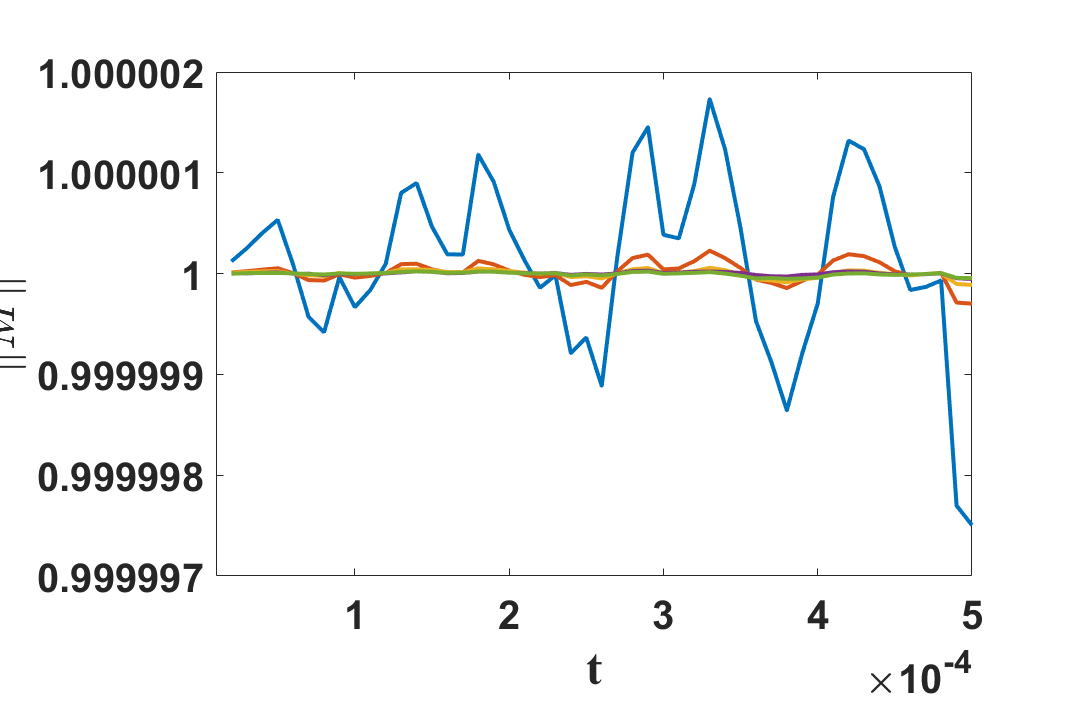}
}
\mbox{
 \hspace{0.45cm}
 \makebox[0.45\textwidth]{(a)}
  \hspace{0.05\textwidth}
 \makebox[0.45\textwidth]{(b)}
 }
\centering 
\mbox{ 
    \includegraphics[width=0.5\textwidth]{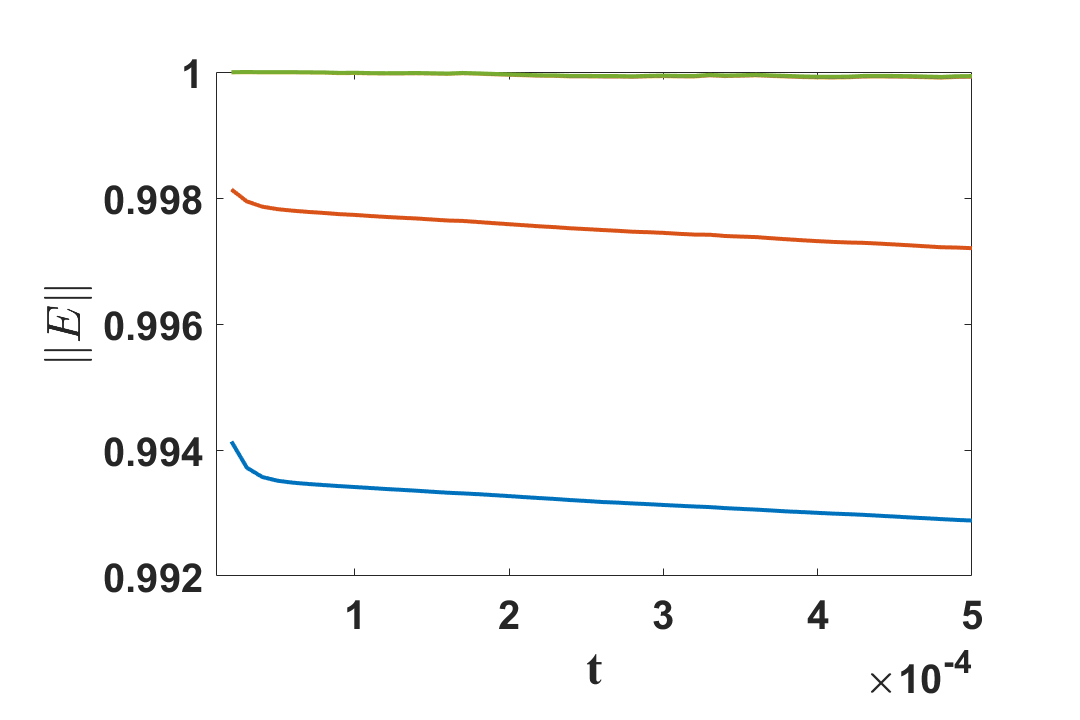}}
\mbox{
 \hspace{0.45cm}
 \makebox[0.45\textwidth]{(c)}
 }
 
\caption{Diffusion of a two dimensional sine function: DSEM-SL method using one element, $N_s$=$10^{4}$ samples and $\Delta t$=$10^{-5}$. The $L^2$ error norm, $\|e\|_{L^2}$, the mass norm, $\|M\|$, and the energy norm, $\|E\|$ are plotted versus time, $t$, in subfigures (a), (b) and (c), respectively.} 
\label{fig:2d_sine_time_evol}
\end{figure}

Figure \ref{fig:2d_sine_time_evol}a plots the time evolution of $\|e\|_{L^2}$ error for $H$=1 and $N_s$=$10^4$ samples. It shows that time evolution of the error decreases as the polynomial order increases for $P$=3 till $P$=7 where the order of the interpolation error becomes equal to the order of the sampling error. Figure \ref{fig:2d_sine_time_evol}b and c plot the time evolution of mass and energy conservation respectively. Conservation is satisfied up-to 3 decimal places for the cases from $P$=3 to $P$=6.   

\begin{figure} 
\centering 
\mbox{ \includegraphics[width=0.5\textwidth]{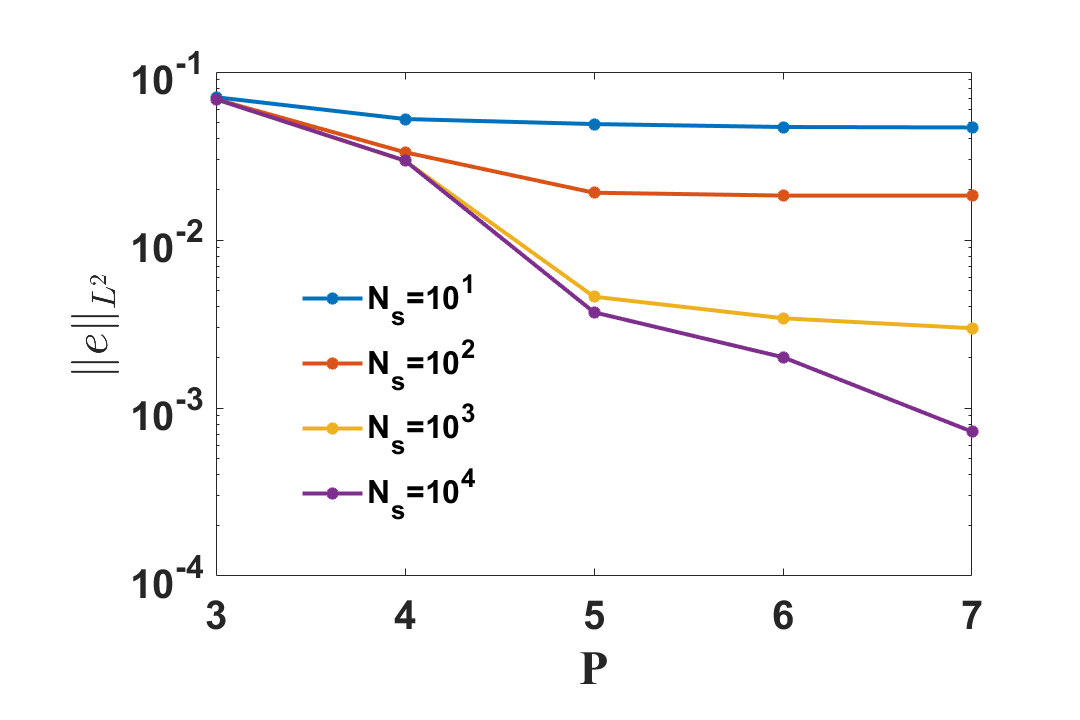}
       \includegraphics[width=0.5\textwidth]{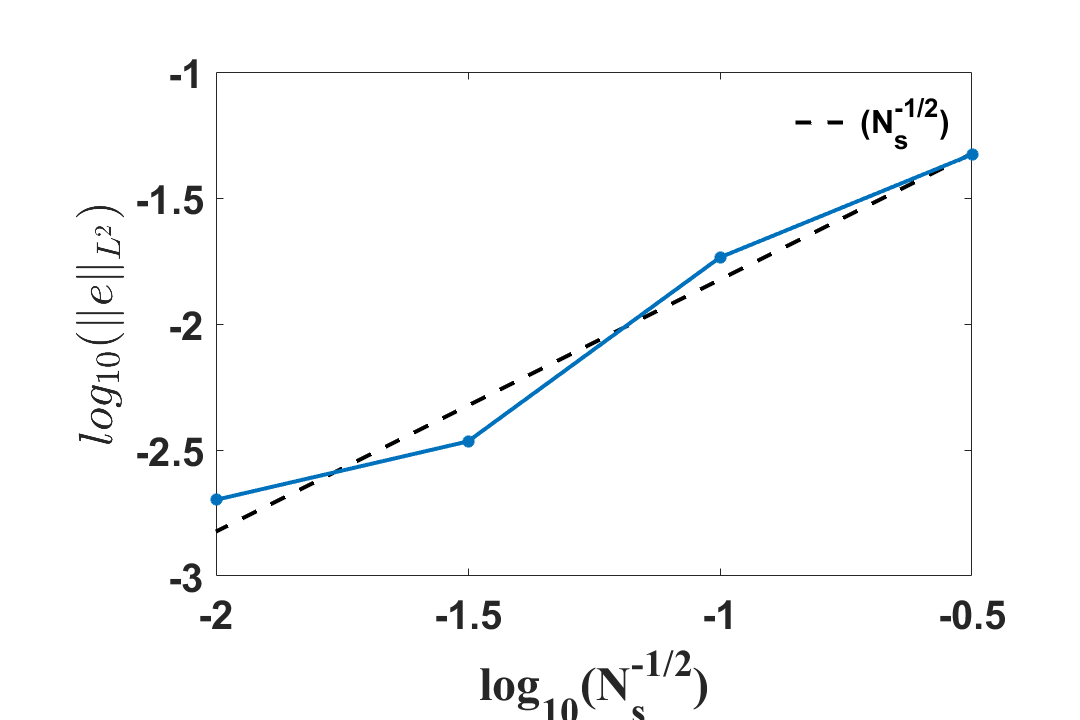}
      }
\mbox{
 \hspace{0.45cm}
 \makebox[0.45\textwidth]{(a)}
 \hspace{0.05\textwidth}
 \makebox[0.45\textwidth]{(b)}
 }
 
\caption{Diffusion of a two dimensional sine wave using DSEM-SL method using one element and $\Delta t$=$10^{-5}$. (a) plots the $P$ convergence of the $L^2$ error for different number of samples, $N_s$. (b) plots the $N_s$ convergence of the error when $P$=7.} 
\label{fig:2d_sine_pconv}
\end{figure}

Figure \ref{fig:2d_sine_pconv}a plots the $P$ convergence of the $\|e\|_{L^2}$ error for the number of samples ranging from, $N_s$= $10$ to $10^4$. The plot shows the expected exponential convergence in $P$ until the $\|e\|_{L^2}$ polynomial error is of the order of the sampling error. Figure \ref{fig:2d_sine_pconv}b shows the linear trend of the sampling error versus $\log_{10}({N_s^{-1/2})}$ which is similar to the one-dimensional case.

\label{sec:2d_sine}
\subsection{Transport of species in two-dimensional Navier-Stokes solutions}
To illustrate that the algorithm works when coupled with a Navier-Stokes solver, we consider the transport of species in an unstable temporally developing shear layer.
To this end, we couple a two-dimensional DSEM-SL solver
to a DSEM Navier-Stokes solver. The fluid flow is governed by the Navier-Stokes equations given by,
\begin{equation} \label{eq:NS}
\frac{\partial Q}{\partial t} + \frac{\partial F_i^a}{\partial x_i} - \frac{\partial F_i^v}{\partial x_i} = 0, 
\end{equation}
where,\begin{align}
Q = \begin{pmatrix}
\rho \\
\rho u_1 \\
\rho u_2 \\
\rho e
\end{pmatrix},       
F_i^a = \begin{pmatrix}
\rho u_i \\
\rho u_1 u_i + P\delta_{i1}\\
\rho u_2 u_i + P\delta_{i2}\\
(\rho e + P) u_i
\end{pmatrix},       
F_i^v = \begin{pmatrix}
0 \\
\sigma_{i1} \\
\sigma_{i2} \\
-q_i + u_k \sigma_{ik}
\end{pmatrix}.    
\end{align}     
The scalar transport of the species, $\phi$ is given by,
\begin{equation}
    \frac{\partial \phi}{\partial t} + {u_i} \frac{\partial \phi}{\partial x_i} = {1\over Re} \frac{\partial^2 \phi}{\partial {x_i}^2}
\end{equation}
The initial shear layer profile for the velocity in the $x$-direction, $u_1$ and the species $\phi$ is set according to the tangent hyperbolic function given by,
\begin{eqnarray}
    u_1(y) &=& {1 \over 2}(1+\tanh{y}) +1, \\
    \phi(y) &=& {1 \over 2}(1+\tanh{y}),
\end{eqnarray}
in a domain $x \in [0,30]$ and $y \in [-15,15]$.
Superimposed on the initial velocity field are perturbation
modes determine from linear-stability analysis of a free shear layer (see for example \cite{NatarajanJacobs17}).
Free stream boundary conditions are applied in the $y$-direction and periodic boundary conditions in the $x$-direction.
The Reynolds number
based on the velocity change of the shear layer and the thickness of the shear layer is $Re$=$10^4$. The parameters used for the simulations are, number of elements in the $x$-direction, $H_x$=$18$ and the number of elements on the $y$-direction, $H_y$=$18$ with polynomial order, $P$=8 and the number of samples, $N_s$=$100$.  
\begin{figure} 
\centering 
\mbox{ 
  \includegraphics[width=0.5\textwidth]{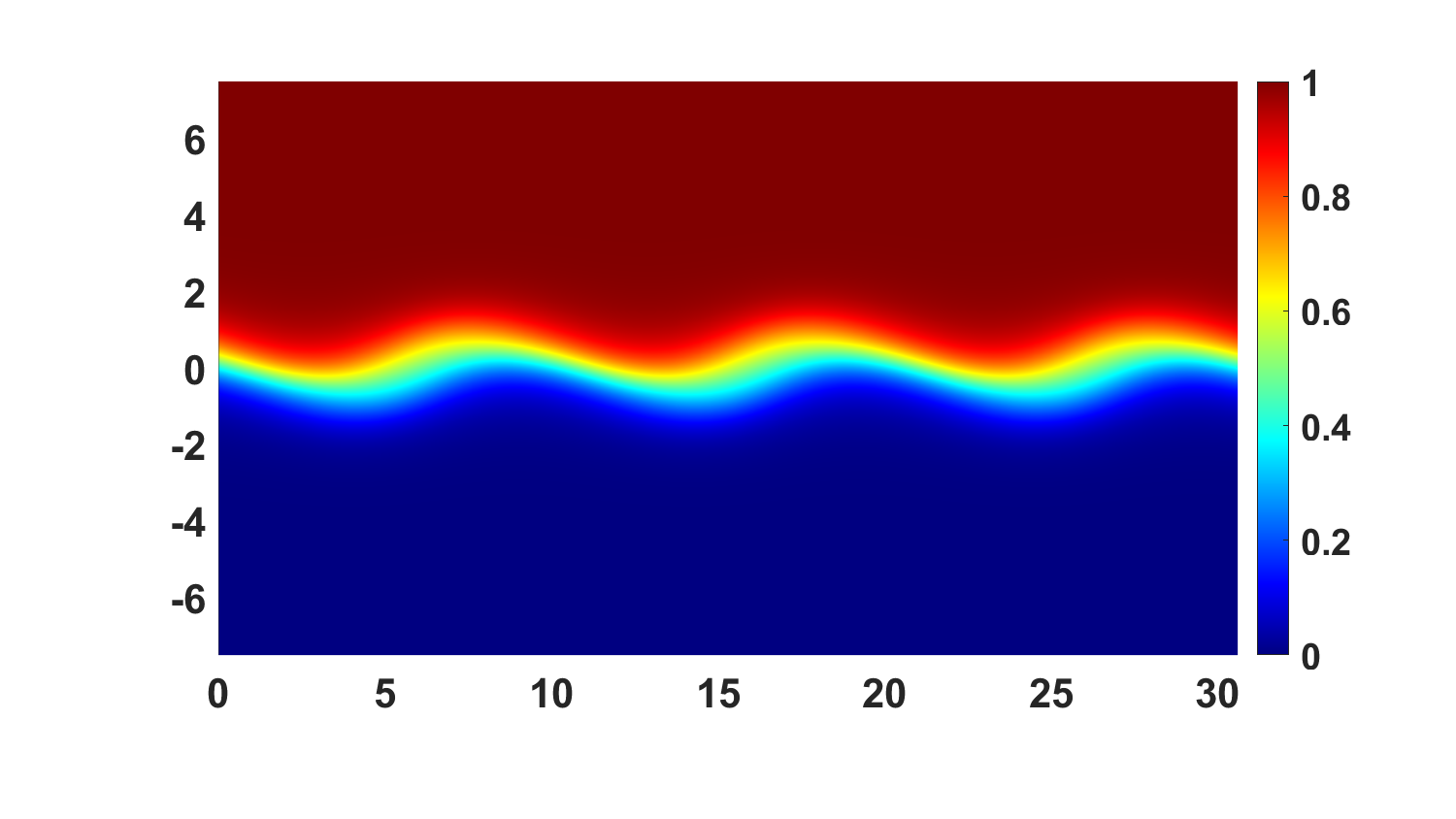}
  \includegraphics[width=0.5\textwidth]{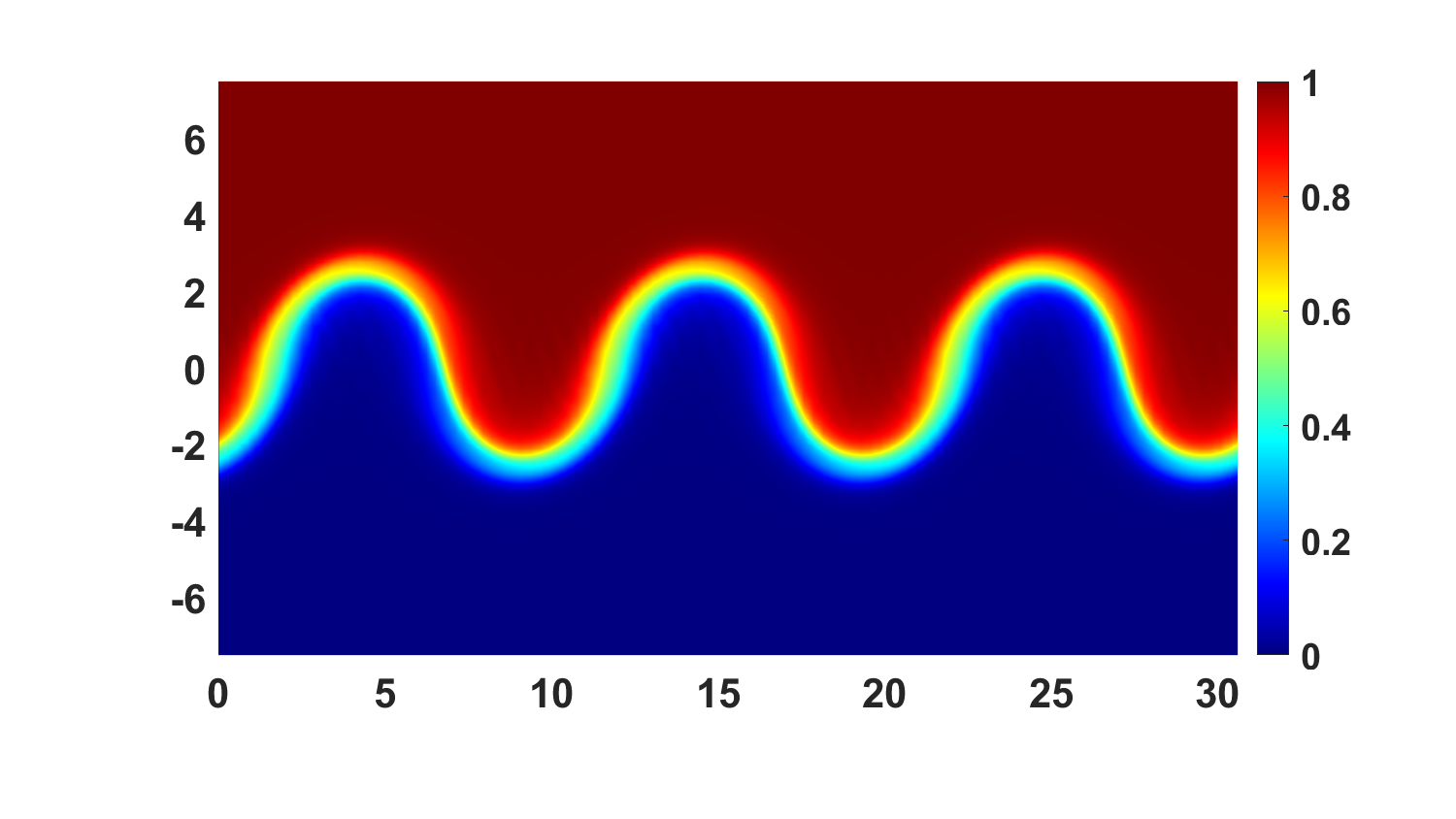}
}
\mbox{
 \hspace{0.45cm}
 \makebox[0.45\textwidth]{(a)}
  \hspace{0.05\textwidth}
 \makebox[0.45\textwidth]{(b)}
 }
\centering 
\mbox{ 
    \includegraphics[width=0.5\textwidth]{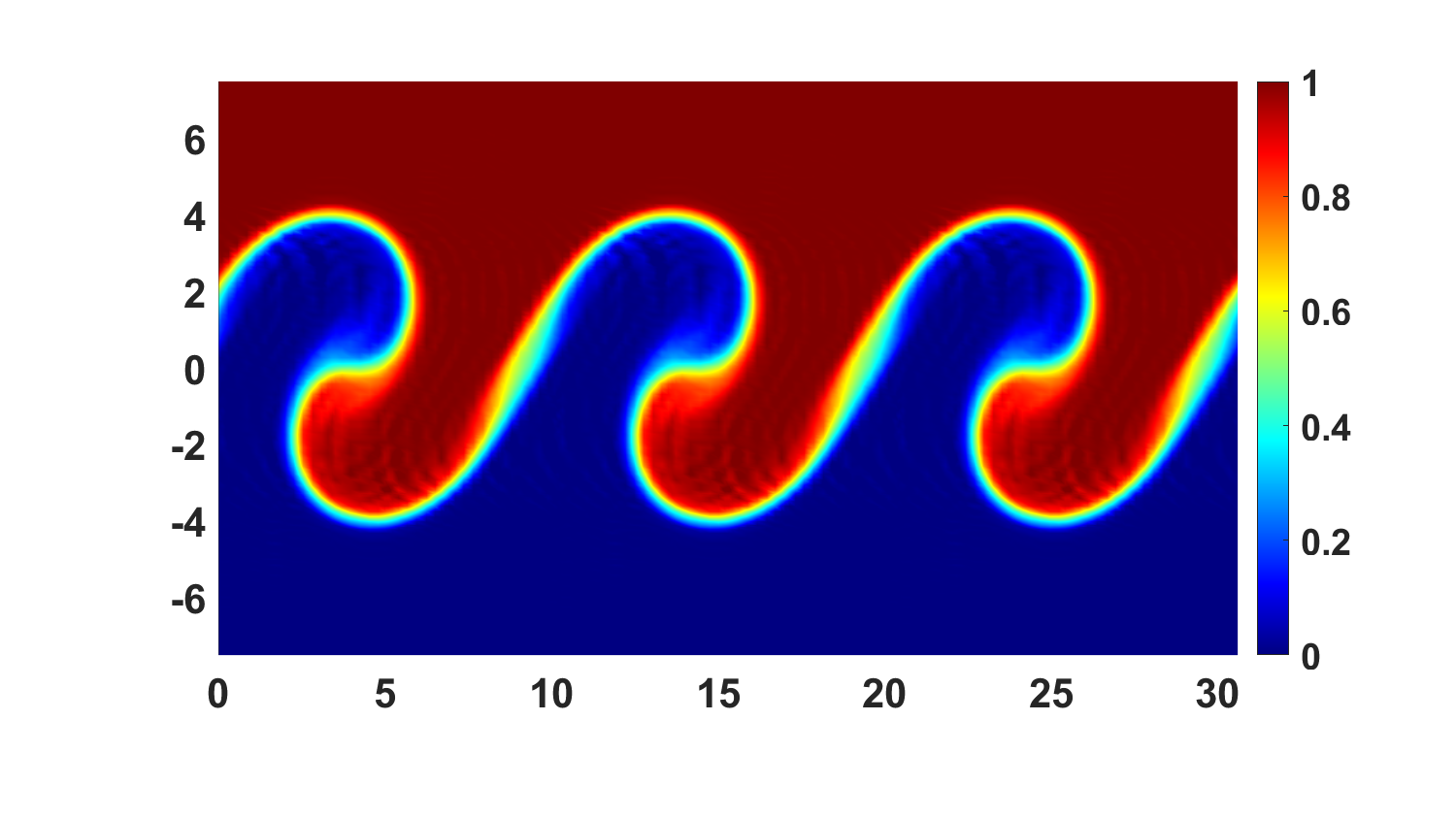}
    \includegraphics[width=0.5\textwidth]{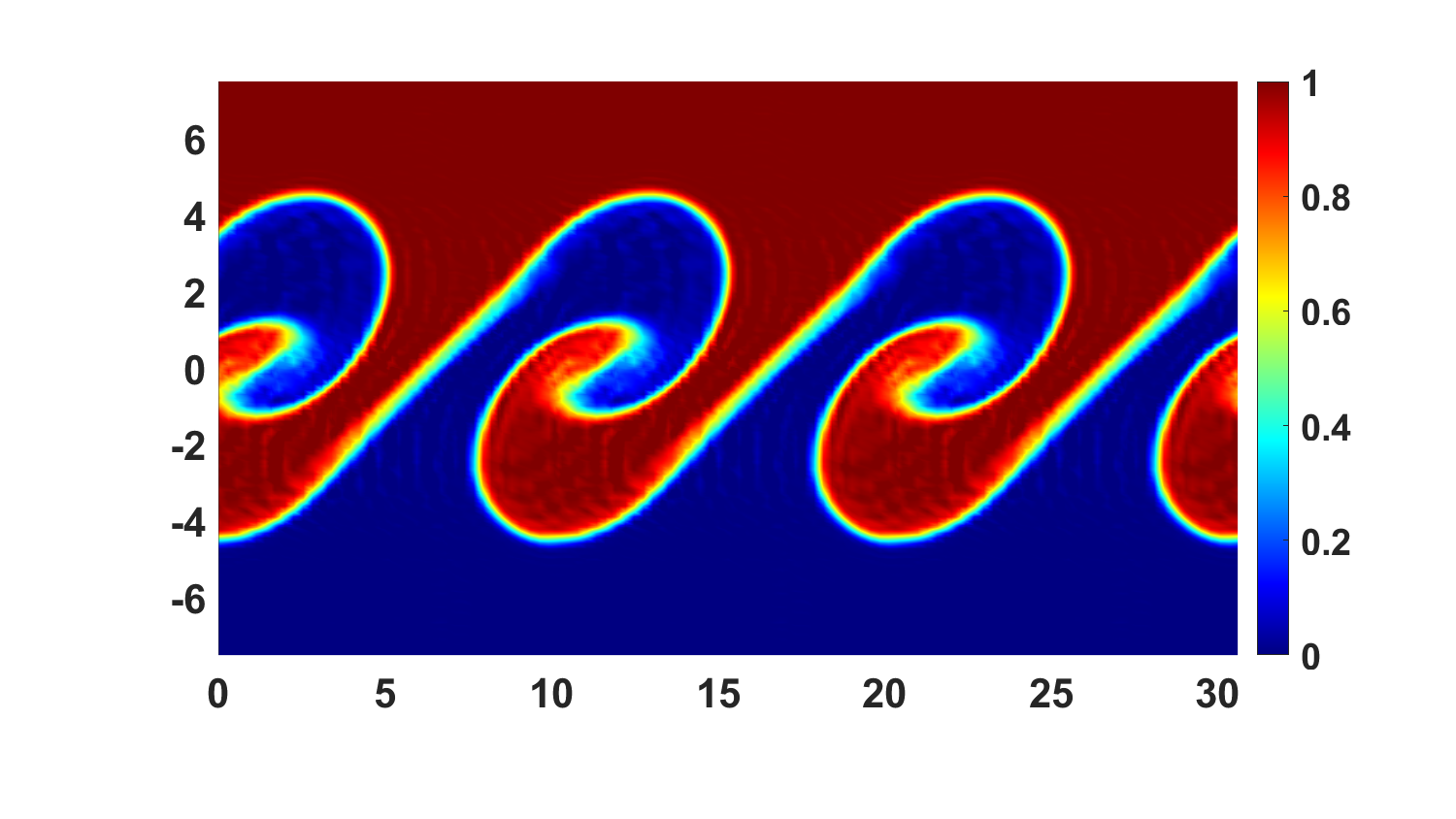}
    }
\mbox{
 \hspace{0.45cm}
 \makebox[0.45\textwidth]{(c)}
  \hspace{0.05\textwidth}
 \makebox[0.45\textwidth]{(d)}
 }
 
\caption{Transport of species in a two-dimensional layer flow configuration. Contour plot of the transport variable, $\phi$ at different time. (a) $t$=1, (b) $t$=5, (c) $t$=10 and (d) $t$=15.  } 
\label{fig:2d_shear_layer}
\end{figure}

Figure \ref{fig:2d_shear_layer} shows the two dimensional contour plot of the transport variable $\phi$ at different time snapshots as determined by the DSEM-SL method. 
Physically, the simulation is of the temporal mixing of two co-flowing species, $A$ and $B$. The transport variable $\phi$ is the species concentration variable with $\phi$=$0$ representing species $A$ and $\phi$=$1$ representing species $B$. The contour plot, Figure \ref{fig:2d_shear_layer}a at time, $t$=$1$ shows the initial linear instability mode developing from the mixing. As the mode develops temporally, in Figures \ref{fig:2d_shear_layer}(b and c), we can observe the non-linear mixing and formation of coherent structures in Figure \ref{fig:2d_shear_layer}d. 
\section{Conclusions}
\label{sec:conclusion}
A semi-Lagrangian method is developed and tested for the consistent and concurrent solution of stochastic Lagrangian differential equations
and Eulerian governing equations approximated with discontinuous spectral element methods (DSEMs). 
The semi-Lagrangian Monte-Carlo approach which accounts for deterministic drift (transport) and 
stochastic diffusion through a Wiener process is proven to be equivalent to solving Eulerian Fokker-Planck type models
for stochastic physics such as filtered density function models for chemically reacting flows. 

The semi-Lagrangian method is consistent with an explicit Eulerian solver
discretized with an explicit DSEM.
By seeding tracer particles at the Gauss quadrature nodes, the Lagrangian
solution is directly available at quadrature nodes of the Eulerian solver and vice-versa. In Eulerian-Lagrangian methods, this exchange of information is commonly performed using computationally intensive and complicated interpolation methods. 

Consistent with DSEM, the semi-Lagrangian method is explicit for the drift term
and uses a Wiener increment for each semi-Lagrangian Monte-Carlo sample. By
choosing the explicit time step and Wiener increment appropriately,  particles are prevented from leaving the element. This ensures a
local and parallel method, which is natural for DSEM. 

Following the explicit trace, the solution is remapped to the original
quadrature points using a least-squares fit.  Element based Monte-Carlo samples are averaged
after the remapping stage at quadrature points only and hence do not require binning and/or distribution functions
with an element as is common the procedure 
for this type of hybrid Eulerian-Lagrangian method. 
For a stable method, it is necessary to update the global solution
according to a single Wiener increment per sample.  Using varying Wiener increments
per quadrature node and/or per element leads to instability in numerical tests.
To prevent steepening of the solution near Dirichlet boundary conditions,
the samples can be periodically  reinitialized with the average of the Monte-Carlo samples.

One-dimensional and two-dimensional tests are conducted for drift-difussion in one and two dimensions,
including for a constant and non-constant diffusion coefficient, and drift-diffusion problems.
The method is shown to be exponentially convergent in space
if the time integration error and sampling or smaller than the spatial approximation error.
Because Monte-Carlo sampling convergence is slow, according to the inverse of the square root of the number of
samples, a significantly smaller number of samples then required for formal spatial convergence is often used.
For a low sampling rate, the semi-Lagrangian method is shown to be stable
and to provide engineering accuracy on the order of a few percent of the solution. 
In a final test the Lagrangian method is coupled with a DSEM based parallel Navier-Stokes solver
and is shown to have optimal parallel performance and provide expected qualitative results.

In current work, we are extending the coupled semi-Lagrangian/Euler solver for simulation
of chemically reacting flow based a filtered density function model as introduced by Givi \cite{Givi99}. 

\section*{Acknowledgements}
Funding provided by the Computational Science Research Center and AFOSR under grant number FA9550-19-1-0387 is greatly appreciated. 

\bibliographystyle{elsarticle-num}
\bibliography{main}

\section*{Appendix A: Proof of consistency}
\label{sec:appendixA}
This Appendix presents a  proof of consistency of the DSEM-SL algorithm introduced in section \ref{sec:dsem_sl},  i.e 
it is shown that in the limit as $N\rightarrow \infty$ and $\Delta t \downarrow 0$, $\phi^s(\mathbf{x},t)$ evolves by a stochastic PDE (\ref{eq:fokkerplanckPhi}).

We start by considering the $s-\mathrm{th}$ sample, and a particle which is initially located at $\mathbf{x}$.
The position of the particle after the advection step is

\begin{equation} \label{eq:proof_dx}
  x^{s*}_j = x_j + u_j \left( \mathbf{x}, t \right)\Delta t + \sqrt{2D\left( \mathbf{x}, t \right)} \Delta W_j^s.
\end{equation}

\noindent Similarly,  according to (\ref{eq:advection_phi}) the value of $\phi^*$ at $\mathbf{x^{s*}}$ after the advection step is

\begin{equation} \label{eq:proof_dphi}
  \phi^{s*}\left( \mathbf{x^{s*}}, t + \Delta t \right) = \phi^{s*} \left( \mathbf{x}, t \right) + S\left( \phi^{s*} \left( \mathbf{x}, t \right); \mathbf{x},t \right)\Delta t,
\end{equation}

\noindent where $S(\psi; \mathbf{x},t)$ is a general source term. The specific version $S(\psi; \mathbf{x},t) = -\psi \frac{\partial u_j}{\partial x_j}$ is used in section \ref{sec:dsem_sl}.

For further analysis,  the variable $\Delta X_j$  representing the difference between the location
before and after advection is introduced as follows:

\begin{equation} \label{eq:deltaX_def}
  \Delta X_j = x^{s*}_j - x_j^{(i)} = u_j \left( \mathbf{x}, t \right)\Delta t + \sqrt{2D\left( \mathbf{x}, t \right)} \Delta W_j^s,
\end{equation}

\noindent Because of the Wiener increment, $\Delta X_j \in O \left(\Delta t ^{1/2} \right)$. 

In this work, the particle is not permitted to leave the bounds of the element, 
and so it follows that $\Delta X_j \in O \left(\Delta x \right)$ with 
$\Delta x$ a representative grid spacing within an element (such as the average 
or minimum grid spacing). 
{The time step thus relates to $\Delta x$  as 
$\Delta t \sim \Delta x^2$ 
} 
and we can Taylor expand in $\mathbf{x}$ from $\phi^{s*}\left( \mathbf{x^{s*}}, t + \Delta t \right)$ to $\phi^{s*}\left( \mathbf{x}, t + \Delta t \right)$, as follows

\begin{eqnarray} \label{eq:proof1}
  \phi^{s*}\left( \mathbf{x}, t + \Delta t \right) & = & \phi^{s*}\left( \mathbf{x^{s*}}, t + \Delta t \right) - \Delta X_j \left. \frac{\partial \phi^{s*}}{\partial x^{s*}_j} \right\vert _{\mathbf{x^{s*}}, t+ \Delta t} + \\
  & &  \frac{1}{2} \Delta X_j \Delta X_j \left. \frac{\partial^2 \phi^{s*}}{\partial x^{s*}_j \partial x^{s*}_j} \right\vert _{\mathbf{x^{s*}}, t + \Delta t} + \littleo(\Delta t) \nonumber,
\end{eqnarray}
\noindent where we use little o notation to denote terms which converge to 0 faster than the little o argument, i.e. $\lim_{y \downarrow 0} \frac{o(y)}{y} = 0$, for any $y$. 
{ The Taylor expansion gives  $\phi^{s*}\left( \mathbf{x}, t + \Delta t \right)$ prior to the remapping step in the semi-Lagagrangian algorithm.
For a sufficiently fine grid, the spectral spatial interpolation error from the remapping step is smaller than $\Delta x^2$ (and therefore $\Delta t$). 
Because the interpolation error of the remapping is $o(\Delta t)$,  $\phi^{s*}\left( \mathbf{x}, t + \Delta t \right)$
after remapping is thus also represented by (\ref{eq:proof1}).}

Substituting (\ref{eq:proof_dphi}) into (\ref{eq:proof1}) we find that
\begin{eqnarray} \label{eq:proof2}
  \phi^{s*}\left( \mathbf{x}, t + \Delta t \right) &=& \phi^{s*}\left( \mathbf{x}, t \right) + S\left( \phi^{s*} \left( \mathbf{x}, t \right); \mathbf{x},t \right)\Delta t - \\
  &-&\Delta X_j \left. \frac{\partial \phi^{s*}}{\partial x^{s*}_j} \right\vert _{\mathbf{x^{s*}},t + \Delta t} + \frac{1}{2} \Delta X_i \Delta X_j \left. \frac{\partial^2 \phi^{s*}}{\partial x^{s*}_i \partial x^{s*}_j} \right\vert _{\mathbf{x^{s*}}, t + \Delta t} + \littleo(\Delta t). \nonumber
\end{eqnarray}

To derive an Eulerian form of this expression, we now seek to  express $\Delta X_j \left. \frac{\partial \phi^{s*}}{\partial x^{s*}_j} \right\vert _{\mathbf{x^{s*}}, t + \Delta t}$ and $\Delta X_i \Delta X_j \left. \frac{\partial^2 \phi^{s*}}{\partial x^{s*}_i \partial x^{s*}_j} \right\vert _{\mathbf{x^{s*}}, t + \Delta t}$ in terms of  $\mathbf{x}$ only. By the chain rule, we have that 

\begin{equation} \label{eq:proof3}
 \left. \frac{\partial \phi^{s*}}{\partial x^{s*}_j} \right\vert _{\mathbf{x^{s*}}, t + \Delta t} = \left. \frac{\partial \phi^{s*}}{\partial x_k} \right\vert _{\mathbf{x}, t + \Delta t}  \left. \frac{\partial x_k}{\partial x^{s*}_j} \right\vert _{\mathbf{x^{s*}}, t + \Delta t}.
\end{equation}

\noindent By (\ref{eq:proof_dphi}), we have that
  
\begin{equation} \label{eq:proof4}
  \left. \frac{\partial \phi^{s*}}{\partial x_k} \right\vert _{\mathbf{x}, t + \Delta t} = \left. \frac{\partial \phi^{s*}}{\partial x_k} \right\vert _{\mathbf{x}, t} + \Delta t \frac{\partial S\left( \phi^{s*} \left( \mathbf{x}, t \right); \mathbf{x},t\right)}{\partial x_k}.
\end{equation}

\noindent By (\ref{eq:proof_dx}) the inverse derivative of $\left. \frac{\partial x_k}{\partial x^{s*}_j} \right\vert _{\mathbf{x^{s*}}, t + \Delta t}$ is 

\begin{equation} \label{eq:proof5}
    \left. \frac{\partial x^{s*}_j}{\partial x_k} \right\vert _{\mathbf{x}, t} = \delta_{jk} + \Delta t \left. \frac{\partial u_j}{\partial x_k} \right\vert _{\mathbf{x},t} + \Delta W_j^s \left. \frac{\partial \left( \sqrt{2D} \right)}{\partial x_k} \right\vert _{\mathbf{x},t},
\end{equation}

\noindent Applying the matrix inverse formula $(I+A)^{-1} = I - A + A^2 + \cdots$ and the identity $\Delta W_j^s \Delta W_k^s = \Delta t \delta _{jk}$  \cite{Evans2013}, we find that 

\begin{equation} \label{eq:proof6}
  \left. \frac{\partial x_k}{\partial x^{s*}_j} \right\vert _{\mathbf{x^{s*}}, t + \Delta t} = \delta_{kj} - \Delta t \left( \frac{\partial u_k}{\partial x_j} - \frac{\partial \sqrt{2D}}{\partial x_k} \frac{\partial \sqrt{2D}}{\partial x_j} \right) - \Delta W_k^s \frac{\partial \sqrt{2D}}{\partial x_j} + \littleo(\Delta t),
\end{equation}

\noindent where all the terms on the right hand side are evaluated at $(\mathbf{x},t)$; we will use the convention for the rest of the derivation that a term is evaluated at $(\mathbf{x},t)$ if there is not specific indication otherwise.  Substituting (\ref{eq:proof6}) and (\ref{eq:proof4}) into (\ref{eq:proof3}) it follows that

\begin{equation}\label{eq:proof7}
  \Delta X_j \left. \frac{\partial \phi^{s*}}{\partial x^{s*}_j} \right\vert _{\mathbf{x^{s*}}, t + \Delta t} = \Delta X_j   \left(  \frac{\partial \phi^{s*}}{\partial x_j} - \frac{\partial \phi^{s*}}{\partial x_k} \Delta W_k^s \frac{\partial \sqrt{2D}}{\partial x_j}  \right) + o(\Delta t).
\end{equation}

Using the same analysis as in (\ref{eq:proof3}-\ref{eq:proof6}), we get that 

\begin{equation}\label{eq:proof8}
  \Delta X_i \Delta X_j \left. \frac{\partial^2 \phi^{s*}}{\partial x^{s*}_i \partial x^{s*}_j} \right\vert _{\mathbf{x^{s*}}, t + \Delta t} = \Delta X_i \Delta X_j \frac{\partial^2 \phi^{s*}}{\partial x_i \partial x_j} + o(\Delta t).
\end{equation}

Substituting (\ref{eq:proof7},\ref{eq:proof8}) into (\ref{eq:proof2}), we can simplify using the identities $\Delta X_j =  u_j \left( \mathbf{x}, t \right)\Delta t + \sqrt{2D\left( \mathbf{x}, t \right)} \Delta W_j^s$ and $\Delta W_j^s \Delta W_k^s = \Delta t \delta _{jk}$ and we get

\begin{eqnarray}\label{eq:proof9}
  \phi^{s*}\left( \mathbf{x}, t + \Delta t \right) - \phi^{s*}\left( \mathbf{x}, t \right) &=& \Delta t \left( S\left( \phi^{s*} \left( \mathbf{x}, t \right); \mathbf{x},t \right) - u_j \frac{\partial \phi^{s*}}{\partial x_j} + \frac{\partial}{\partial x_j}\left( D \frac{\partial \phi ^{s*}}{\partial x_j} \right) \right) - \nonumber \\
  &-&\frac{\partial \phi^{s*}}{\partial x_j} \sqrt{2D} \Delta W_j^s + o(\Delta t).
\end{eqnarray}

Finally, dividing (\ref{eq:proof9}) by $\Delta t$ and taking the limit as $\Delta t \downarrow 0$, we get that 

\begin{equation}\label{eq:proof10}
  \frac{\partial \phi^{s*}}{\partial t} + u_j \frac{\partial \phi^{s*}}{\partial x_j} + \frac{\partial \phi^{s*}}{\partial x_j} \sqrt{2D} \dot{W}_j^s - \frac{\partial}{\partial x_j}\left( D \frac{\partial \phi ^{s*}}{\partial x_j}\right) = S\left( \phi^{s*}; \mathbf{x},t \right),
\end{equation}

\noindent where $\dot{W}_j^s$ is the weak time derivative of the multivariate Wiener process. This is a stochastic PDE (in Ito form) which uniquely defines the PDF $\Pdf_\phi(\psi;\mathbf{x},t)$ of the random variable $\phi^*$. Note that, just as $\Delta W_j^s$ is the same Wiener increment for all initial points $\mathbf{x}$ in the $s-\mathrm{th}$ sample (i.e., all collocation points), then so also is $\dot{W}_j^s$ spatially independent. The above derivations shed light on why this spatial independence is necessary, since the derivative $\left. \frac{\partial x_k}{\partial x^{s*}_j} \right\vert _{\mathbf{x^{s*}}, t + \Delta t}$ from (\ref{eq:proof6}) would not be well-defined if the Wiener increments for different points $\mathbf{x}$ in the $s-\mathrm{th}$ sample were independent.

From (\ref{eq:proof10}), we can derive the Fokker-Planck equation for $\Pdf_\phi(\psi;\mathbf{x},t)$. To do this, we use the following Lemma from \cite{Sabelnikov05}:

\begin{lemma}[Sabel'nikov and Soulard]\label{lemma1}
If the stochastic field $\phi^{s*}(\mathbf{x},t)$ evolves by the stochastic PDE (in Stratanovich form)

\begin{equation}\label{eq:proof11}
\frac{\partial \phi^{s*}}{\partial t}dt +dv_j \circ \frac{\partial \phi^{s*}}{\partial x_j} = S(\phi^{s*};\mathbf{x},t),
\end{equation}

\noindent with $dv_j$ being a stochastic advection term, and $S(\psi;\mathbf{x},t)$ being a deterministic and Lipschitz continuous function of $\psi,\mathbf{x}$ and $t$, then the Fokker-Planck equation for the PDF, $\Pdf_\phi(\psi;\mathbf{x},t)$, of $\phi^{s*}$ is

\begin{equation}\label{eq:proof12}
\frac{\partial \Pdf_\phi}{\partial t} + \left( \frac{\left\langle dv_j \right\rangle}{dt} +\frac{1}{2} \frac{ \left\langle dv_j \frac{\partial dv_k}{\partial x_k} \right\rangle}{dt} \right) \frac{\partial \Pdf_\phi}{\partial x_j} = \frac{\partial}{\partial x_j}\left( \frac{1}{2} \frac{\left\langle dv_j dv_k \right\rangle}{dt} \frac{\partial \Pdf_\phi}{\partial x_k} \right) - \frac{\partial}{\partial \psi}\left[ S(\psi;\mathbf{x},t) \Pdf_\phi \right].
\end{equation}
\end{lemma}

To use Lemma \ref{lemma1}, we need to cast (\ref{eq:proof10}) in Stratanovich form. From Evans \cite{Evans2013} we have the following Ito to Stratanovich conversion formula:

\begin{equation}\label{eq:proof13}
\mathbf{B}\left(\mathbf{W},t\right)\circ d\mathbf{W} = \mathbf{B}\left(\mathbf{W},t\right) d\mathbf{W} + \frac{1}{2} \frac{\partial B_{ij}}{\partial x_j}\left(\mathbf{W},t\right)dt,
\end{equation}

\noindent where $\mathbf{W}$ is the multivariate Wiener process, and $\mathbf{B}(\mathbf{y},t)$ is a $C^1$ matrix function, with $B_{ij}$ being its components. Applying (\ref{eq:proof13}) to $B_{ij}\left( \mathbf{W},t \right) = \sqrt{2D(\mathbf{x},t)} \delta_{i1} \frac{\partial \phi^{s*}}{\partial x_j}$ for a fixed $\mathbf{x}$, we have that

\begin{eqnarray}\label{eq:proof14}
 \sqrt{2D(\mathbf{x},t)} dW_j \circ \frac{\partial \phi^{s*}}{\partial x_j} &=& \sqrt{2D(\mathbf{x},t)} \frac{\partial \phi^{s*}}{\partial x_j} \circ dW_j \nonumber \\
&\mkern-180mu=\mkern+180mu&\mkern-180mu \sqrt{2D(\mathbf{x},t)} \frac{\partial \phi^{s*}}{\partial x_j} dW_j + \frac{1}{2}\frac{\partial}{\partial W_j}\left( \sqrt{2D(\mathbf{x},t)} \frac{\partial \phi^{s*}}{\partial x_j} \right) dt \nonumber \\
&\mkern-180mu=\mkern+180mu&\mkern-180mu \sqrt{2D(\mathbf{x},t)} \frac{\partial \phi^{s*}}{\partial x_j} dW_j + \frac{\sqrt{2D(\mathbf{x},t)}}{2}\frac{\partial}{\partial x_j}\left( \frac{\partial \phi^{s*}}{\partial W_j} \right) dt \nonumber \\
&\mkern-180mu=\mkern+180mu&\mkern-180mu \sqrt{2D(\mathbf{x},t)} \frac{\partial \phi^{s*}}{\partial x_j} dW_j - \frac{\sqrt{2D(\mathbf{x},t)}}{2}\frac{\partial}{\partial x_j}\left( \sqrt{2D(\mathbf{x},t)} \frac{\partial \phi^{s*}}{\partial x_j} \right) dt\nonumber \\
&\mkern-180mu=\mkern+180mu&\mkern-180mu \sqrt{2D} \frac{\partial \phi^{s*}}{\partial x_j} dW_j - \frac{\partial}{\partial x_j}\left( D \frac{\partial \phi^{s*}}{\partial x_j} \right) dt+ \sqrt{D} \frac{\partial \sqrt{D}}{\partial x_j} \frac{\partial \phi^{s*}}{\partial x_j}  dt
\end{eqnarray}

\noindent where we used  that $\sqrt{2D(\mathbf{x},t)}$ is independent of $\mathbf{W}$ for the equality on the first line 
and to get from the second to the third line.
To get from the first to the second line, we applied (\ref{eq:proof10})
 (\ref{eq:proof13}) is finally used   to get from the third to the fourth line. Multiplying (\ref{eq:proof10}) by $dt$ and substituting the last line of (\ref{eq:proof14}) into it, we find that

\begin{equation}\label{eq:proof15}
  \frac{\partial \phi^{s*}}{\partial t} dt + u_j \frac{\partial \phi^{s*}}{\partial x_j} dt +  \sqrt{2D} d{W}_j^s \circ \frac{\partial \phi^{s*}}{\partial x_j} - \sqrt{D} \frac{\partial \sqrt{D}}{\partial x_j} \frac{\partial \phi^{s*}}{\partial x_j}dt  = S\left( \phi^{s*}; \mathbf{x},t \right) dt.
\end{equation}

\noindent With  $dv_j \equiv \left( u_j - \sqrt{D} \frac{\partial \sqrt{D}}{\partial x_j} \right) dt + \sqrt{2D} d{W}_j^s$, we have that

\begin{equation}\label{eq:proof16}
\frac{ \left\langle dv_j \right\rangle}{dt} = \frac{\left( u_j - \sqrt{D} \frac{\partial \sqrt{D}}{\partial x_j} \right) dt}{dt} = u_j - \sqrt{D} \frac{\partial \sqrt{D}}{\partial x_j},
\end{equation}

\noindent since $d{W}_j^s$ has zero mean. Similarly, because $dv_j = \sqrt{2D} d{W}_j^s + o(dt^{1/2})$ and $\frac{\partial dv_k}{\partial x_k} = \frac{\partial \sqrt{2D}}{\partial x_k} d{W}_j^s + o(dt^{1/2})$, and $\left\langle dW_i dW_j \right\rangle = \delta_{ij} dt$, the second-order mean terms in (\ref{eq:proof12}) are respectively

\begin{equation}\label{eq:proof17}
\frac{1}{2} \frac{ \left\langle dv_j \frac{\partial dv_k}{\partial x_k} \right\rangle}{dt} = \sqrt{D} \frac{\partial \sqrt{D}}{\partial x_j},
\end{equation}

and 

\begin{equation}\label{eq:proof18}
\frac{1}{2} \frac{\left\langle dv_j dv_k \right\rangle}{dt} = D\delta_{jk}.
\end{equation}

Substituting (\ref{eq:proof16}-\ref{eq:proof18}) into (\ref{eq:proof12}) and simplifying, we get the Fokker-Planck equation for $\Pdf(\psi;\mathbf{x},t)$

\begin{equation}\label{fokkerPlanckProof}
\frac{\partial \Pdf_\phi}{\partial t} + u_j \frac{\partial \Pdf_\phi}{\partial x_j} = \frac{\partial}{\partial x_j} \left( D \frac{\partial \Pdf_\phi}{\partial x_j} \right) - \frac{\partial}{\partial \psi}\left[ S(\psi;\mathbf{x},t) \Pdf_\phi \right],
\end{equation}

\noindent which is identical to (\ref{eq:fokkerplanckPhi})

\end{document}